\documentclass[%
 aip,
 jmp,%
 amsmath,amssymb,
 preprint,
]{revtex4-1}

\usepackage{graphicx}
\usepackage{dcolumn}
\usepackage{bm}

\renewcommand*{\Re}{\operatorname{Re}}
\renewcommand*{\Im}{\operatorname{Im}}

\begin{document}

\title{Variational theory of the tapered impedance transformer}

\author{R. P. Erickson}
\affiliation{Self-Energy LLC, Scottsdale, Arizona 85259, USA}
\email{r.p.erickson@icloud.com}

\date{\today}

\begin{abstract}
Superconducting amplifiers are key components of modern quantum information circuits. To minimize information loss and reduce oscillations a tapered impedance transformer of new design is needed at the input/output for compliance with other 50 $\Omega$ components. We show that an optimal tapered transformer of length $\ell$, joining amplifier to input line, can be constructed using a variational principle applied to the linearized Riccati equation describing the voltage reflection coefficient of the taper. For an incident signal of frequency $\omega_o$ the variational solution results in an infinite set of equivalent optimal transformers, each with the same form for the reflection coefficient, each able to eliminate input-line reflections. For the special case of optimal lossless transformers, the group velocity $v_g$ is shown to be constant, with characteristic impedance dependent on frequency $\omega_c=\pi v_g / \ell$. While these solutions inhibit input-line reflections only for frequency $\omega_o$, a subset of optimal lossless transformers with $\omega_o$ significantly detuned from $\omega_c$ does exhibit a wide bandpass. Specifically, by choosing $\omega_o\rightarrow 0$ ($\omega_o\rightarrow\infty$), we obtain a subset of optimal low-pass (high-pass) lossless tapers with bandwidth $(0,\sim\omega_c)$ ($(\sim\omega_c,\infty)$). From the subset of solutions we derive both the wide-band low-pass and high-pass transformers, and we discuss the extent to which they can be realized given fabrication constraints. In particular, we demonstrate the superior reflection response of our high-pass transformer when compared to other taper designs. Our results have application to amplifier, transceiver, and other components sensitive to impedance mismatch.
\end{abstract}

\maketitle

\section{\label{sec:introduction}Introduction}
Tapered impedance transformers are ubiquitous and represent important components in the design of microwave to millimeter transmissions lines, low-noise amplifiers, and transceivers. They address the impedance mismatch between input and load-bearing lines that tends to induce undesirable signal reflections, power loss, and poor signal-to-noise characteristics. Over the years different types of taper designs have been presented that reduce the size of reflections for MHz frequencies and greater. The most notable ones are discussed in standard engineering texts, such as Pozar.\cite{Pozar2012} 

For example, the Klopfenstein taper\cite{Klopfenstein1956} is a particularly popular high-pass design used extensively in modern high-speed electronics because the maximum-reflection parameter of the model can be set below the threshold of reflection sensitivity of the application. A typical parameter setting is a maximum reflection of 2\% within the range of frequencies of the passband.\footnote{Specifically, in the Klopfenstein model of Ref~(\onlinecite{Klopfenstein1956}), as it is discussed in Ref.~(\onlinecite{Pozar2012}), by a maximum reflection of 2\% in the passband we mean the maximum reflection parameter $\Gamma_m$ is set to $\Gamma_m=0.02$.} Since, in many cases, a reflection coefficient of 5\% to 10\% is tolerable, the Klopfenstein taper is more than adequate. In a few instances, such as superconducting nonlinear parametric amplifiers,\cite{CastellanosBeltran2008,Bergreal2010,Spietz2010,Hatridge2011,SLUG,EomNature2012amplifier,Roch2012,Mutas2013,Macklin2015,Vissers2016} with added noise the order of 1 photon, small reflections of a few percent are readily amplified when the device is operated at a pump frequency of about $1-10$ GHz, which can lead to poor signal-to-noise output.

In particular, superconducting amplifiers are a key component of modern quantum information circuits, and represent the motivation for the present study. In order to obtain high performance, e.g., quantum-limited noise and wide bandwidth, it is not always possible to maintain a 50 $\Omega$ environment due to high kinetic inductance\cite{EomNature2012amplifier,Vissers2016,Erickson2017} and Josephson junction capacitances.\cite{Mutas2013,Macklin2015,MohebbiPhD2010thesis,Chaudhuri2015} In order to minimize information loss and reduce oscillations in the circuit it is therefore necessary to use a tapered impedance transformer on the input and/or output of these devices in order to be compliant with other 50 $\Omega$ components. Due to absence of loss, these circuits are challenging because non-ideal behavior such as small reflections can quickly build up and cause undesirable oscillations and sharp frequency-dependent response. This is particularly applicable to the case of traveling-wave amplifiers,\cite{EomNature2012amplifier,Vissers2016,Erickson2017} which require extremely wide bandwidth and smooth response to support multiple idlers and various high-frequency pumps.

Use of the Klopfenstein and other high-pass tapers to address impedance mismatch in these instances is not ideal. For example, in the case of the Klopfenstein taper the maximum ripple within the passband is designed to be constant, but cannot be made sufficiently small to reduce corresponding ripple in the signal gain of the traveling-wave amplifier, whereas tapers like the triangular and exponential designs described in Pozar\cite{Pozar2012} actually perform somewhat better in this regard.\footnote{D. P. Pappas, private communication.} Presumably this is because these latter designs exhibit asymptotic drop-off of ripple across the exploitable passband; while the reflection drop-off is no better than $1/\omega^2$ with increasing frequency $\omega$, it is sufficient to enable these latter designs to outperform the Klopfenstein taper at the higher pump and idler frequencies encountered in the traveling-wave amplifier. Furthermore, the claim that any of these aforementioned tapers are optimal is not rigorously justified from a mathematical standpoint. An optimal solution must be determined via comparison with all other reasonable possibilities, which therefore suggests that these tapers can be improved upon.

In the construction of an optimal impedance taper the input-line reflection coefficient is the measurable quantity of interest that must be engineered to zero. In fact, as we show in Appendix~\ref{appendix:DifferentialEquation}, a discontinuity exists between the zero reflection coefficient of the input line and the reflection coefficient just inside the taper. In the early work of Collin\cite{Collin1956} a high-pass taper was derived from an $N$-section quarter-wave cascaded transformer structure by taking the continuum limit of $N\rightarrow\infty$. In the Klopfenstein taper design, the characteristic impedance of the taper was deduced from the Fourier transform of the reflection coefficient just inside the taper, which in turn was formed from an ansatz consistent with the results of Collin.\cite{Klopfenstein1956} In both of these earlier treatments the assumption is that an optimal high-pass taper may be constructed via a procedure that minimizes reflections at every cross section along the length of the taper. A better approach is to treat the minimization of the input-line reflections via a variational principle, wherein the optimal reflection coefficient as a function of position along the length of the taper follows from the variational procedure itself. This later approach implicitly compares taper profiles and selects only those that are truly optimal with respect to input-line reflections.

In the discussion that follows, we apply a variational approach to obtain a mathematically accurate definition of the optimal impedance transformer for the general case of an input line connected to a load-bearing transmission line. We have in mind a waveguide in place of the transmission line, but our method is also valid for the case when there is a terminated load after the tapered transformer. Specifically, for a signal of frequency $\omega_o$ incident to the transformer, we vary the magnitude of the input-line voltage reflections to obtain the form of the taper corresponding to the absolute minimum of reflections, i.e., zero input-line reflections. This is accomplished without a priori assumption about continuity of the reflection coefficient across the input-line/taper interface. We show that there is an infinite number of equivalent tapers, each with its own reflection coefficient within the transformer, which share the common property of having zero reflections in the input line, specifically for the input frequency $\omega_o$. Because this set of equivalent solutions arises from the absolute minimum of a variation principle, it is therefore justified to refer to each as an optimal impedance transformer for signals of frequency $\omega_o$.

One problem with an optimal impedance transformer as defined above is that, in general, it is only applicable to the specific frequency $\omega_o$ for which it is designed. This is a consequence of the Bode-Fano criterion,\cite{Bode1945,*Fano1950-1} which prevents perfectly zero reflections over an extended range of frequencies. Nevertheless, any impedance transformer design must have a significant bandpass to be of practical use. To resolve this narrow-bandwidth issue we calculate the reflection response along the input line for a signal of arbitrary frequency $\omega$ incident upon an optimal lossless transformer of design frequency $\omega_o$ obtained from the variational principle.

By construction the input-line reflection response will be precisely zero only when $\omega=\omega_o$. However, as we show, the reflection response is dependent on a characteristic frequency $\omega_c=\pi v_g/\ell$, where $v_g$ is the constant transformer group velocity and $\ell$ is the transformer length. Only when $\omega\cong\omega_c$, for which the wavelength of the incident signal is about $2\ell$, does the magnitude of the reflection response along the input line become larger. When a transformer design is considered for which $\omega_o$ is significantly detuned from $\omega_c$ then the magnitude of the reflection response becomes very small, over an extended range of frequencies $\omega$, provided $\omega$ is closer to $\omega_o$ than $\omega_c$.

Thus, an optimal wide-bandwidth lossless impedance transformer is an optimal transformer whose design frequency $\omega_o$ is significantly detuned from its characteristic frequency $\omega_c$. Moreover, if we take the limit $\omega_o\rightarrow 0$ of the transformer design then the detuning establishes a bandpass $0<\omega\lesssim\omega_c$, corresponding to a low-pass transformer. Conversely, if we take the limit $\omega_o\rightarrow\infty$ then detuning implies a bandpass $\omega_c\lesssim\omega<\infty$, corresponding to a high-pass transformer. In this way, an optimal wide-bandwidth impedance transformer, of either low-pass or high-pass character, is obtained from an optimal transformer design by taking the appropriate limit of the design frequency $\omega_o$, which is as far from $\omega_c$ as possible.

In what follows we apply our variational approach to obtain the reflection coefficient, propagation coefficient, and characteristic impedance of a set of optimal impedance transformers, assuming an incident signal of frequency $\omega_o$. For arbitrary frequency $\omega$ we then calculate the input-line reflection response of an optimal lossless transformer of design frequency $\omega_o$. By detuning design frequency $\omega_o$ from characteristic frequency $\omega_c$, we derive the reflection response and characteristic impedance of both the low-pass ($\omega_o\rightarrow 0$) and high-pass ($\omega_o\rightarrow\infty$) cases. In each case, by examining the asymptotic behavior of the reflection response as $\omega\rightarrow\omega_o$, we show how to obtain a lossless transformer with negligible reflections over a wide bandpass. In particular, for the high-pass case, we compare our solution to other transformer designs.\cite{Pozar2012} We also discuss the extent to which both of our solutions can be realized given fabrication constraints. 
 
\section{\label{sec:theory}The Variational Theory}
\begin{figure}
\includegraphics[width=240pt, height=152pt]{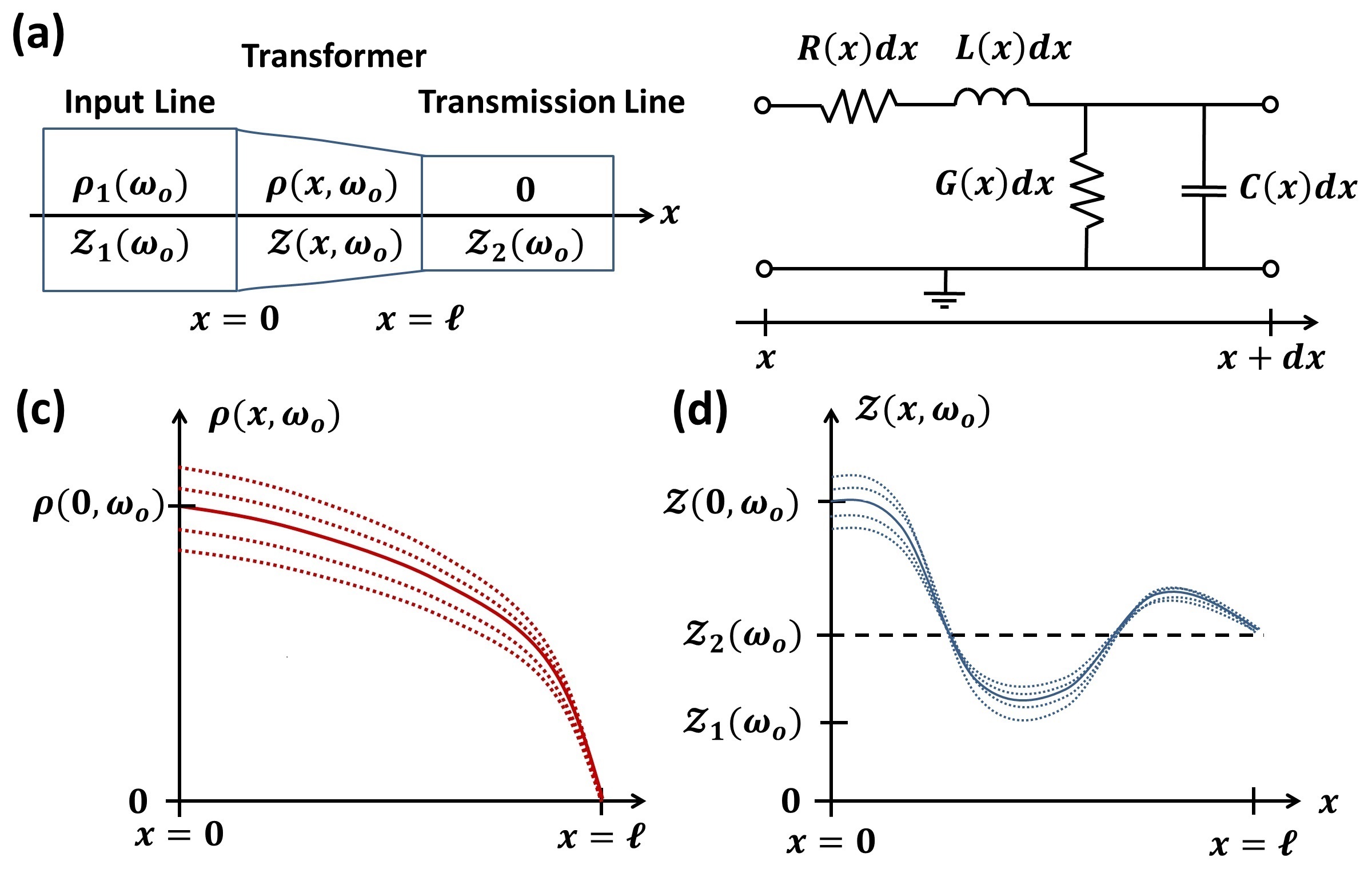}
\caption{\label{fig1} (a) Sketch of impedance transformer of length $\ell$, of characteristic impedance $\mathcal{Z}(x,\omega_o)$ and reflection coefficient $\rho(x,\omega_o)$ at each point $0<x<\ell$, placed between input line and load-bearing transmission line of characteristic impedances $\mathcal{Z}_1(\omega_o)$ and $\mathcal{Z}_2(\omega_o)$, respectively. The transformer is designed to minimize input-line reflection coefficient $\rho_1(\omega_o)$. (b) Depiction of infinitesimal rung of length $dx$, at $x$, of a ladder-type transmission-line model of the components of (a). Also, we illustrate hypothetical solutions of: (c) $\rho(x,\omega_o)$ and (d) $\mathcal{Z}(x,\omega_o)$, as a function of $x$. Solid curves represent optimal solutions that render $|\rho_1(\omega_o)|$ minimal.}
\end{figure}

We consider an input line with forward-traveling wave of frequency $\omega_o$ incident upon a transmission line possessing a tapered interval $0\le x\le \ell$, as in Fig.~\ref{fig1}(a). The boundary between input and transmission lines is at $x=0$. The characteristic impedance of the input line is $\mathcal{Z}_1(\omega_o)$, whereas in the tapered interval it is $\mathcal{Z}(x,\omega_o)$. As $x\rightarrow\ell$, $\mathcal{Z}(x,\omega_o)$ smoothly transitions to $\mathcal{Z}_2(\omega_o)$ of the transmission-line interior, i.e., $\mathcal{Z}(\ell,\omega_o)=\mathcal{Z}_2(\omega_o)$. The measurable reflection coefficient of the traveling wave within the input line is $\rho_1(\omega_o)$ while inside the taper it is $\rho(x,\omega_o)$. 

In Appendix \ref{appendix:DifferentialEquation} we derive the Riccati differential equation of Walker and Wax\cite{Walker1946} satisfied by $\rho(x,\omega_o)$; importantly, we include the accompanying boundary conditions. Assuming ${|\rho(x,\omega_o)|}^2 \ll 1$, the equation may be expressed in linearized form as
\begin{equation} \label{eq:diffeq}
\frac{\partial}{\partial x} \rho(x,\omega_o) \cong
2 \gamma(x,\omega_o) \rho(x,\omega_o) - \frac{1}{2} \frac{\partial}{\partial x} \log{\mathcal{Z}(x,\omega_o)} ; \;\;
{|\rho(x,\omega_o)|}^2 \ll 1 ,
\end{equation}
with $\gamma(x,\omega_o)$ representing the propagation coefficient. From Eqs.~(\ref{eq:rho-1}) and (\ref{eq:rho-ell}), the accompanying boundary conditions at $x=0$ and $x=\ell$ are, respectively,
\begin{subequations} \label{eq:rho-bcs}
\begin{eqnarray}
\rho_1(\omega_o) = \frac{ \mathcal{Z}(0,\omega_o) \left[ 1 + \rho(0,\omega_o) \right] - \mathcal{Z}_1 \left[ 1 - \rho(0,\omega_o) \right] }
{ \mathcal{Z}(0,\omega_o) \left[ 1 + \rho(0,\omega_o) \right] + \mathcal{Z}_1 \left[ 1 - \rho(0,\omega_o) \right] } ,  \label{eq:rho-bcs-1} \\
\rho(\ell,\omega_o)=0 .  \label{eq:rho-bcs-2}
\end{eqnarray}
\end{subequations}

To determine the optimal tapered impedance transformer we minimize $|\rho_1(\omega_o)|$. Ultimately, we want to set $|\rho_1(\omega_o)|=0$, but important information can be obtained through a formal minimization. From Eq.~(\ref{eq:rho-bcs-1}) we see that $|\rho_1(\omega_o)|$ can be reduced if a proper choice of $\mathcal{Z}(0,\omega_o)$ and $\rho(0,\omega_o)$ is made. This is possible because, unlike boundary conditions $\mathcal{Z}(\ell,\omega_o)=\mathcal{Z}_2(\omega_o)$ and $\rho(\ell,\omega_o)=0$, the boundary values of $\mathcal{Z}(0,\omega_o)$ and $\rho(0,\omega_o)$ are not firmly established. 

In Eq.~(\ref{eq:diffeq}), $\rho(x,\omega_o)$ depends on both $\gamma(x,\omega_o)$ and $\mathcal{Z}(x,\omega_o)$, which may be expressed as
\begin{subequations} \label{eq:gamma-zee-naught}
\begin{eqnarray}
\gamma(x,\omega_o)=\sqrt{Z(x,\omega_o)Y(x,\omega_o)} , \label{eq:gamma-zee-naught-1} \\
\mathcal{Z}(x,\omega_o)=\sqrt{Z(x,\omega_o)/Y(x,\omega_o)} , \label{eq:gamma-zee-naught-2}
\end{eqnarray}
\end{subequations}
respectively, where $Z(x,\omega_o)=R(x)+i\omega_o L(x)$ is the series impedance per unit length and $Y(x,\omega_o)=G(x)+i\omega_o C(x)$ is the shunt admittance per unit length. As described in greater detail in Appendix \ref{appendix:DifferentialEquation}, $R(x)$, $L(x)$, $G(x)$, and $C(x)$ are the unit-length resistance, inductance, conductance, and capacitance, respectively, of the tapered region, as introduced via the ladder-type transmission-line model depicted in Fig.~\ref{fig1}(b). 

Our goal of minimizing $|\rho_1(\omega_o)|$ may be described as adjusting the underlying values of independent variables $Z(x,\omega_o)$ and $Y(x,\omega_o)$ at each point $x$ of the taper, subject to fixed boundary conditions of $\rho(\ell,\omega_o)=0$ and $\mathcal{Z}(\ell,\omega_o)=\mathcal{Z}_2(\omega_o)$, such that the values of $\gamma(x,\omega_o)$ and $\mathcal{Z}(x,\omega_o)$, and thus also $\rho(x,\omega_o)$, may be altered to give boundary values $\rho(0,\omega_o)$ and $\mathcal{Z}(0,\omega_o)$ that make $|\rho_1(\omega_o)|$ as small as possible. Panels (c) and (d) of Fig.~\ref{fig1} depict this idea, where the various curves of $\rho(x,\omega_o)$ and $\mathcal{Z}(x,\omega_o)$ are generated as $Z(x,\omega_o)$ and $Y(x,\omega_o)$ are varied. The solid curves have corresponding boundary values for $\rho(0,\omega_o)$ and $\mathcal{Z}(0,\omega_o)$ that render $|\rho_1(\omega_o)|$ minimal. The underlying $Z(x,\omega_o)$ and $Y(x,\omega_o)$ of these solid curves define the optimal impedance transformer for the incident traveling wave of frequency $\omega_o$.

To optimize $|\rho_1(\omega_o)|$ in the manner of variational calculus,\cite{Mathews1970} we vary both underlying functions $Z(x,\omega_o)$ and $Y(x,\omega_o)$, holding endpoint $x=\ell$ fixed but allowing endpoint $x=0$ to float. Specifically, we let $Z(x,\omega_o)\rightarrow Z(x,\omega_o) + \delta Z(x,\omega_o)$ and $Y(x,\omega_o)\rightarrow Y(x,\omega_o) + \delta Y(x,\omega_o)$ where $\delta Z(\ell,\omega_o)=0$ and $\delta Y(\ell,\omega_o)=0$ but $\delta Z(0,\omega_o)\ne 0$ and $\delta Y(0,\omega_o)\ne 0$. Since the variation of $|\rho_1(\omega_o)|$ is equivalent to varying $\rho_1(\omega_o)$, we vary $\rho_1(\omega_o)$ such that from Eq.~(\ref{eq:rho-bcs-1}) we have 
\begin{multline} \label{eq:variation-1}
\delta \rho_1(\omega_o) \cong 
\mathcal{Z}_1(\omega_o) \mathcal{Z}(0,\omega_o)  \big[ 4 \, \delta \rho(0,\omega_o) 
+ \delta Z(0,\omega_o) \left/ Z(0,\omega_o) \right. - \delta Y(0,\omega_o) \left/ Y(0,\omega_o) \right. \big] \\
\times { \left\{ \mathcal{Z}(0,\omega_o) \left[ 1 + \rho(0,\omega_o) \right] + \mathcal{Z}_1(\omega_o) 
\left[ 1 - \rho(0,\omega_o) \right] \right\} }^{-2}  ; \;\;
{|\rho(0,\omega_o)|}^2 \ll 1 .
\end{multline}

To obtain $\delta\rho(0,\omega_o)$ in Eq.~(\ref{eq:variation-1}), we first integrate Eq.~(\ref{eq:diffeq}) so that $\rho(0,\omega_o)$ may be expressed as a functional of $Z(x,\omega_o)$, $Y(x,\omega_o)$ and their derivatives, viz.
\begin{equation} \label{eq:rho-end}
\rho(0,\omega_o)[Z,Z';Y,Y'] = 
\int_0^\ell \Bigg[ -2 \rho(x,\omega_o) \sqrt{Z(x,\omega_o) Y(x,\omega_o)} 
+ \frac{Z'(x,\omega_o)}{4Z(x,\omega_o)} 
- \frac{Y'(x,\omega_o)}{4Y(x,\omega_o)} \Bigg] dx ,
\end{equation}
where $\rho(x,\omega_o)$ is implicitly a function of $Z(x,\omega_o)$, $Y(x,\omega_o)$ and derivatives. Then, using Eq.~(\ref{eq:rho-end}) to obtain $\delta\rho(0,\omega_o)$, and subsequently setting $\delta\rho_1(\omega_o)=0$ in Eq.~(\ref{eq:variation-1}), we arrive at two Euler-Lagrange equations that may be expressed as
\begin{equation} \label{eq:euler-lagrange-Z}
\sqrt{Y/Z} \, \rho + 2 \sqrt{Z Y} \, \frac{\partial\rho}{\partial Z} - 2 \frac{d}{d x} \left[ \sqrt{Z Y} \frac{\partial\rho}{\partial Z'} \right] = 0 , 
\end{equation}
\begin{equation} \label{eq:euler-lagrange-Y}
\sqrt{Z/Y} \, \rho + 2 \sqrt{Z Y} \, \frac{\partial\rho}{\partial Y} - 2 \frac{d}{d x} \left[ \sqrt{Z Y} \frac{\partial\rho}{\partial Y'} \right] = 0 , 
\end{equation}
with boundary conditions at $x=0$ given by
\begin{equation} \label{eq:boundary-conditions}
{ \left[ \sqrt{Z Y} \frac{\partial\rho}{\partial Z'} \right] }_{x=0} = 0 , \;\;
{ \left[ \sqrt{Z Y} \frac{\partial\rho}{\partial Y'} \right] }_{x=0} = 0 .
\end{equation}

Normally, one would solve Eqs.~(\ref{eq:euler-lagrange-Z}) through (\ref{eq:boundary-conditions}) for $Z(x,\omega_o)$ and $Y(x,\omega_o)$. However, since $\rho(x,\omega_o)$ is unknown, a more fruitful approach is to solve for $\rho(x,\omega_o)$ in terms of $Z(x,\omega_o)$ and $Y(x,\omega_o)$. This gives
\begin{equation}
\rho(x,\omega_o) = \left[ \gamma(0,\omega_o) / \gamma(x,\omega_o) \right] 
\left[ A(x,\omega_o) + B(\omega_o) Z'(x,\omega_o) + C(\omega_o) Y'(x,\omega_o) \right] , 
\end{equation}
where $A(x,\omega_o)$ is an arbitrary function of $x$ and $B(\omega_o)$ and $C(\omega_o)$ are independent of $x$. If we apply this form to the boundary conditions of Eq.~(\ref{eq:boundary-conditions}) we further see that $B(\omega_o)=0$ and $C(\omega_o)=0$. Thus, the reflection coefficient within the optimal transformer is of the form $\rho(x,\omega_o) = \left[ \gamma(0,\omega_o) / \gamma(x,\omega_o) \right] A(x,\omega_o)$. 

The arbitrariness of $A(x,\omega_o)$ allows us immediately to set $|\rho_1(\omega_o)|=0$, or equivalently, $\rho_1(\omega_o)=0$. In this way, our optimization always results in the absolute minimum, $|\rho_1(\omega_o)|=0$. Specifically, we set $\rho_1(\omega_o)=0$ in Eq.~(\ref{eq:rho-bcs-1}) and solve for $\rho(0,\omega_o)$, obtaining a second boundary condition on $\rho(x,\omega_o)$, viz.
\begin{equation} \label{eq:perfect}
\rho(0,\omega_o) = \frac{ \mathcal{Z}_1(\omega_o) - \mathcal{Z}(0,\omega_o) }{ \mathcal{Z}_1(\omega_o) + \mathcal{Z}(0,\omega_o) } ,
\end{equation}
in addition to $\rho(\ell,\omega_o)=0$. Then, rescaling $A(x,\omega_o)$ by letting $A(x,\omega_o)=\rho(0,\omega_o) f(x,\omega_o)$, where $f(x,\omega_o)$ is now the arbitrary function of $x$, and incorporating the additional boundary condition of Eq.~(\ref{eq:perfect}), we now have
\begin{equation} \label{eq:rho-solution}
\rho(x,\omega_o) = \left[ \frac{ \mathcal{Z}_1(\omega_o) - \mathcal{Z}(0,\omega_o) }{ \mathcal{Z}_1(\omega_o) + \mathcal{Z}(0,\omega_o) } \right] 
\left[ \frac{ \gamma(0,\omega_o) }{ \gamma(x,\omega_o) } \right] f(x,\omega_o); \;\;
f(0,\omega_o) = 1, \; f(\ell,\omega_o) = 0 ,
\end{equation}
where imposition of $f(0,\omega_o)=1$ and $f(\ell,\omega_o)=0$ ensures the boundary conditions of $\rho(x,\omega_o)$ are satisfied. Assuming a physically meaningful solution for $Z(x,\omega_o)$ and $Y(x,\omega_o)$ exists, Eq.~(\ref{eq:rho-solution}) is the optimal form of $\rho(x,\omega_o)$ for any impedance transformer that eliminates input-line reflections, specifically for frequency $\omega_o$. 

\begin{widetext}
Given the optimal form of $\rho(x,\omega_o)$, we may determine the optimal taper design by substituting Eq.~(\ref{eq:rho-solution}) into Eq.~(\ref{eq:diffeq}). This yields the constraint
\begin{multline} \label{eq:constraint-0}
2 \gamma(0,\omega_o) \left[ \mathcal{Z}_1(\omega_o) - \mathcal{Z}(0,\omega_o) \right] f(x,\omega_o) \\
+ \gamma(x,\omega_o) \left\{ 4 \gamma(0,\omega_o) \left[ \mathcal{Z}_1(\omega_o) - \mathcal{Z}(0,\omega_o) \right] \int_x^{\ell} f(x') \, dx' 
+ \left[ \mathcal{Z}_1(\omega_o) + \mathcal{Z}(0,\omega_o) \right] \log{\frac{\mathcal{Z}(x,\omega_o)}{\mathcal{Z}_2(\omega_o)}} \right\} = 0 .
\end{multline}
\end{widetext}
This equation determines optimal $\gamma(x,\omega_o)$ and $\mathcal{Z}(x,\omega_o)$, or equivalently, via Eqs.~(\ref{eq:gamma-zee-naught}), optimal $Z(x,\omega_o)$ and $Y(x,\omega_o)$.\footnote{Equation~(\ref{eq:constraint-0}) involves functions of complex numbers, so it actually represents two real-valued equations. For an optimal lossless transformer this is sufficient to determine the characteristic impedance and the propagation coefficient because the former is a real number while the later is an imaginary number. However, for the case of an optimal lossy taper, it may not be possible to specify the optimal characteristic impedance and/or propagation coefficient definitively, unless additional information about the nature of the lossiness, i.e., the transmission-line series resistance and shunt conductance, is also provided.} Since $f(x,\omega_o)$ is an arbitrary function of $x$, subject to $f(0,\omega_o)=1$ and $f(\ell,\omega_o)=0$, there are an infinite number of optimal transformer designs, where each design is characterized by $\omega_o$ and $f(x,\omega_o)$.\footnote{Here $f(x,\omega_o)$ resembles a gauge transformation on scalar field $\rho(x,\omega_o)$ where $|\rho_1(\omega_o)|=0$ is invariant.}

Several key points of our variational approach are:
\begin{enumerate}
\item The boundary condition of Eq.~(\ref{eq:rho-bcs-1}) illustrates the discontinuity of the voltage reflection coefficient across the $x=0$ interface.
\item A variation of the input line reflection coefficient, $\rho_1(\omega_o)$, with endpoint $x=0$ not fixed, is developed from Eq.~(\ref{eq:rho-bcs-1}), applicable to a specific frequency $\omega_o$ of incident forward-traveling wave.
\item The variation of $\rho_1(\omega_o)$ employs the variation of $Z(x,\omega_o)$ and $Y(x,\omega_o)$ at each point $x$ along the length of the taper, which necessitates the variation of $\rho(x,\omega_o)$, the voltage reflection coefficient within the taper, since it depends on both $Z(x,\omega_o)$ and $Y(x,\omega_o)$.
\item The optimization of $\rho_1(\omega_o)$ implies an optimal form for $\rho(x,\omega_o)$, as given by Eq.~(\ref{eq:rho-solution}), where $f(x,\omega_o)$ is subject to the stated boundary conditions; any other proposed form for $\rho(x,\omega_o)$ will not correspond to an optimal $\rho_1(\omega_o)$. 
\item The optimal form of $Z(x,\omega_o)$ and $Y(x,\omega_o)$, or equivalently $\gamma(x,\omega_o)$ and $\mathcal{Z}(x,\omega_o)$, are obtained from Eq.~(\ref{eq:constraint-0}).
\end{enumerate}
In what follows we narrow our discussion to the case of a lossless optimal impedance transformer.

\section{\label{sec:lossless}The Optimal Lossless Impedance Transformer}
For the remainder of our discussion we assume input line, transformer, and transmission line are lossless, such that
\begin{subequations} \label{eq:lossless}
\begin{eqnarray}
\mathcal{Z}_1 = \sqrt{L_1/C_1} , \;\; \mathcal{Z}_2 = \sqrt{L_2/C_2} , \label{eq:lossless-1} \\
\mathcal{Z}(x,\omega_o) = \sqrt{L(x,\omega_o) \left/ C(x,\omega_o) \right.} , \label{eq:lossless-2} \\ 
\gamma(x,\omega_o)=i\omega_o \sqrt{L(x,\omega_o) C(x,\omega_o)} , \label{eq:lossless-3}
\end{eqnarray}
\end{subequations}
where $L(x,\omega_o)$ and $C(x,\omega_o)$ are, respectively, the transformer inductance and capacitance per unit length at $x$. Also, $L_1$, $C_1$ and $L_2$, $C_2$ are the constant values of the input line and interior of the transmission line, respectively, with $L(\ell,\omega_o)=L_2$ and $C(\ell,\omega_o)=C_2$. 

In Appendix~\ref{appendix:LosslessSolution} we apply Eqs.~(\ref{eq:lossless}) to Eq.~(\ref{eq:constraint-0}) to obtain the solution of the optimal lossless impedance transformer. An important result, which follows from Eq.~(\ref{eq:choice-for-f}) and Eq.~(\ref{eq:g-bc}), is that $f(x,\omega_o)$ of an optimal lossless transformer may be written in the form
\begin{equation} \label{eq:f}
f(x,\omega_o) = \frac{d}{dx} g(x) + 2\pi i \left( \frac{\omega_o}{\omega_c} \right) g(x) \left/ \ell \right. ,
\end{equation}
where $\omega_c=\pi/\left( \ell \sqrt{L_2 C_2} \right)$ is a frequency characteristic of the impedance transformer, and $g(x)$ is a real-valued function of $x$ satisfying the boundary conditions
\begin{equation} \label{eq:g}
g(0) = 0 , \;\;\ g(\ell) = 0 , \;\; { \left. \frac{d}{dx} g(x) \right| }_{x=0} = 1 , \;\; { \left. \frac{d}{dx} g(x) \right| }_{x=\ell} = 0 .
\end{equation}
In this way, any design of optimal lossless impedance transformer is characterized by both $\omega_o$ and $g(x)$.

Also in our analysis of Appendix \ref{appendix:LosslessSolution}, via Eq.~(\ref{eq:LC-solution}), the propagation coefficient $\gamma(x,\omega_o)=i\omega_o\sqrt{L(x)C(x)}$ of the lossless transformer is found to be a constant value, i.e., $\gamma(x,\omega_o)=i\omega_o\sqrt{L_2C_2}$. In the literature\cite{Klopfenstein1956} one typically approaches the problem of finding an optimal impedance transformer by assuming $\gamma(x,\omega_o)$ is independent of $x$. Here, using our variational approach, the optimal propagation coefficient is indeed independent of $x$, the same value as in the interior of the transmission line. 

Since the dispersion frequency of the transformer is $\Omega(k)=k/\sqrt{L_2C_2}$, where $k$ is a wavenumber, the group velocity is $v_g=\partial\Omega(k)/\partial k=1/\sqrt{L_2C_2}$, which is constant for any $g(x)$, allowing us to write $\omega_c=\pi v_g/\ell$. Similarly, for an incident signal of frequency $\omega=\Omega(k)$, the phase velocity is $v_p=\omega/k=v_g$. Thus, we may interpret $\tau_c=2\pi/\omega_c=2\ell/v_g$ as the time it takes for a signal of frequency $\omega=\omega_c$ to traverse the length of the transformer and be reflected back to the input-line/taper interface. In this case the signal wavelength is $\lambda=v_p \tau_c=2\ell$ since $v_p=v_g$.

Via Eq.~(\ref{eq:char-imp-1}), $\mathcal{Z}(x,\omega_o)$ of the optimal lossless transformer may be expressed in terms of $g(x)$ as
\begin{widetext}
\begin{equation} \label{eq:Z}
\mathcal{Z}(x,\omega_o) = \mathcal{Z}_2 \exp{ \left\{ \left[ \log{ \frac{ \mathcal{Z}(0,\omega_o) }{ \mathcal{Z}_2 } } \right] 
\left[ \frac{ d g(x) \left/ dx \right. - { \left( 2\pi\omega_o \left/ \omega_c \right. \right) }^2 \int_x^{\ell} g(x') \left/ \ell^2 \right. dx' }
{ 1 - { \left( 2\pi\omega_o \left/ \omega_c \right. \right) }^2 \int_0^{\ell} g(x') \left/ \ell^2 \right. dx' } \right] \right\} } ,
\end{equation}
where, via Eq.~(\ref{eq:solution-0}), $\mathcal{Z}(0,\omega_o)$ is a root of
\begin{equation} \label{eq:Z0}
2 \left[ \mathcal{Z}_1 - \mathcal{Z}(0,\omega_o) \right] 
\left[ 1 - { \left( 2\pi\omega_o \left/ \omega_c \right. \right) }^2 \frac{1}{\ell^2} \int_0^{\ell} g(x') \, dx' \right] 
+ \left[ \mathcal{Z}_1 + \mathcal{Z}(0,\omega_o) \right] \log{ \frac{\mathcal{Z}(0,\omega_o)}{\mathcal{Z}_2} } = 0 .
\end{equation}
\end{widetext}
Since $\gamma(x,\omega_o)=i\omega_o\sqrt{L_2C_2}$ is independent of $x$, the inductance per unit length and the capacitance per unit length can be obtained from $L(x,\omega_o) = \sqrt{L_2 C_2} \; \mathcal{Z}(x,\omega_o)$ and $C(x,\omega_o) = \sqrt{L_2 C_2} / \mathcal{Z}(x,\omega_o)$, respectively.

We summarize the key points of the optimal lossless transformer as follows:
\begin{enumerate}
\item For the optimal lossless transformer $f(x,\omega_o)$ takes the particular form of Eq.~(\ref{eq:f}), as shown in Appendix~\ref{appendix:LosslessSolution}, where $g(x)$ is a real-valued function satisfying the boundary conditions of Eq.~(\ref{eq:g}).
\item Therefore a general result is that the design of the optimal lossless transformer is defined by the choice of $g(x)$ and $\omega_o$.
\item The frequency $\omega_c=\pi v_g/\ell$ is characteristic of the geometry and material composition of the optimal lossless taper, and is therefore more or less fixed, save for some ability to change the geometry, such as via the transformer length $\ell$.
\item An important result of our variational approach applied to the lossless transformer is that the optimal propagation coefficient of this case, $\gamma(x,\omega_o)$, is a constant in $x$, i.e., $\gamma(x,\omega_o)=i\omega_o\sqrt{L_2C_2}$, as demonstrated in Appendix~\ref{appendix:LosslessSolution}. 
\item The optimal characteristic impedance of the lossless taper, $\mathcal{Z}(x,\omega_o)$, is given by Eq.~(\ref{eq:Z}), where the boundary value $\mathcal{Z}(0,\omega_o)$ is determined from the transcendental Eq.~(\ref{eq:Z0}).
\end{enumerate}

\subsection{\label{subsection:response}Reflection Response of the Optimal Lossless Impedance Transformer}
As mentioned earlier, the optimal lossless impedance transformer of Eqs.~(\ref{eq:Z}) and (\ref{eq:Z0}) guarantees zero input-line reflections only for an incident signal corresponding to frequency $\omega_o$. To determine reflection-response characteristics of the transformer at any other frequency $\omega$ we first solve Eq.~(\ref{eq:diffeq}) for frequency $\omega$, instead of frequency $\omega_o$, but with characteristic impedance given by Eq.~(\ref{eq:Z}). The result is the reflection coefficient of the transformer with respect to $\omega$, which after some algebra may be expressed as
\begin{multline} \label{eq:rho-w}
\rho(x;\omega,\omega_o) = \rho(x,\omega_o) 
+ 2 \pi^2 \left[ \log{ \frac{\mathcal{Z}(0,\omega_o)}{\mathcal{Z}_2} } \right] \\
\times \left( \frac{{\omega_o}^2 - {\omega}^2}{{\omega_c}^2} \right)
\frac{ \int_x^{\ell} g(x') e^{2\pi i \left( \omega \left/ \omega_c \right. \right) \left( x - x' \right) \left/ \ell \right. } \left/ \ell^2 \right. dx' }
{ 1 - { \left( 2\pi\omega_o \left/ \omega_c \right. \right) }^2 \int_0^{\ell} g(x') \left/ \ell^2 \right. dx' } ,
\end{multline}
where
\begin{equation} \label{eq:rho-o-design}
\rho(x,\omega_o) = \left[ \frac{ \mathcal{Z}_1 - \mathcal{Z}(0,\omega_o) }{ \mathcal{Z}_1 + \mathcal{Z}(0,\omega_o) } \right] 
\left[ \frac{d}{dx} g(x) + 2\pi i \left( \frac{\omega_o}{\omega_c} \right) g(x) / \ell \right] .
\end{equation}
Equation (\ref{eq:rho-o-design}) is just the lossless limit of the reflection coefficient of Eq.~(\ref{eq:rho-solution}), with $f(x,\omega_o)$ given by Eq.~(\ref{eq:f}) and $\gamma(x,\omega_o)$ independent of $x$. 

For arbitrary $\omega$ the input-line reflection coefficient is still of the form of Eq.~(\ref{eq:rho-bcs-1}), but now
\begin{equation} \label{eq:rho1-w-temp}
\rho_1(\omega,\omega_o) = 
\frac{ \mathcal{Z}(0,\omega_o) \left[ 1 + \rho(0;\omega,\omega_o) \right] - \mathcal{Z}_1 \left[ 1 - \rho(0;\omega,\omega_o) \right] }
{ \mathcal{Z}(0,\omega_o) \left[ 1 + \rho(0;\omega,\omega_o) \right] + \mathcal{Z}_1 \left[ 1 - \rho(0;\omega,\omega_o) \right] } ,
\end{equation}
where $\rho(0;\omega,\omega_o)$ is Eq.~(\ref{eq:rho-w}) at $x=0$. Also, setting $x=0$ in Eq.~(\ref{eq:rho-o-design}) yields $\rho(0,\omega_o)$, the same as Eq.~(\ref{eq:perfect}). Thus, substituting the $x=0$ form of Eq.~(\ref{eq:rho-w}) into Eq.~(\ref{eq:rho1-w-temp}), and making use of Eq.~(\ref{eq:perfect}) for $\rho(0,\omega_o)$, we may write $\rho_1(\omega,\omega_o)=[\rho(0;\omega,\omega_o)-\rho(0,\omega_o)]/[1-\rho(0;\omega_o,\omega_o)\rho(0,\omega_o)]$. Assuming $|\rho(0;\omega,\omega_o)\rho(0,\omega_o)|\ll 1$, this may be approximated as $\rho_1(\omega,\omega_o)\cong\rho(0;\omega,\omega_o)-\rho(0,\omega_o)$, or from Eq.~(\ref{eq:rho-w}), we have
\begin{equation} \label{eq:rho1-w}
\rho_1(\omega,\omega_o) \cong
2 \pi^2 \left[ \log{ \frac{\mathcal{Z}(0,\omega_o)}{\mathcal{Z}_2} } \right] \left( \frac{{\omega_o}^2 - {\omega}^2}{{\omega_c}^2} \right) 
\frac{ \int_0^{\ell} g(x) e^{-2\pi i \left( \omega \left/ \omega_c \right. \right) x \left/ \ell \right. } \left/ \ell^2 \right. dx }
{ 1 - { \left( 2\pi\omega_o \left/ \omega_c \right. \right) }^2 \int_0^{\ell} g(x) \left/ \ell^2 \right. dx } .
\end{equation}
This is the small-reflection response of a traveling wave of frequency $\omega$ incident upon an optimal impedance transformer of design defined by $\omega_o$ and $g(x)$. 

Equation~(\ref{eq:rho1-w}) may be used to analyze the passband characteristics of any optimal lossless impedance transformer with characteristic impedance of form given by Eqs~(\ref{eq:Z}) and (\ref{eq:Z0}). As examples, we next consider the two wide-band cases delineated by the transformer characteristic frequency $\omega_c$.\footnote{Another case of interest is that of setting the design frequency to $\omega_o=\omega_c$ in Eq.~(\ref{eq:rho1-w}). This eliminates large reflection response attributable to $\omega_c$ and results in a V-like passband centered about frequency $\omega_o$. We consider the definition of this type of notch filter to be a subject for follow-on research.} Several important points the reader should keep in mind as we investigate these cases are:
\begin{enumerate}
\item The optimal lossless transformer of design choice $g(x)$ and $\omega_o$ has reflection coefficient within the taper, $\rho(x,\omega_o)$, given by Eq.~(\ref{eq:rho-o-design}).
\item We may use the reflection-response function, $\rho_1(\omega,\omega_o)$ of Eq.~(\ref{eq:rho1-w}), to determine choices for $g(x)$ and $\omega_o$ that exhibit specific wide-bandwidth characteristics.
\item By construction $\rho_1(\omega_o,\omega_o)=0$ in Eq.~(\ref{eq:rho1-w}), and this is the only point of the $\rho_1(\omega,\omega_o)$ versus $\omega$ curve where $\rho_1(\omega,\omega_o)$ is precisely zero. 
\end{enumerate}

\subsection{\label{subsection:highpass}The Wide-Band High-Pass Lossless Impedance Transformer}
Recall from our introductory remarks that a wide-band high-pass impedance transformer can be constructed if the design frequency $\omega_o$ is detuned from the characteristic frequency $\omega_c$ such that $\omega_o\rightarrow\infty$. This statement is made more explicit by examination of the reflection response $\rho_1(\omega,\omega_o)$ of Eq.~(\ref{eq:rho1-w}). First, note that $\rho_1(\omega_o,\omega_o)=0$ by construction; this is always the case, no matter the value of $\omega_o$. If we let $\omega_o\rightarrow\infty$ then we have
\begin{equation} \label{eq:rho1-hp}
\rho^{(HP)}_1(\omega) = \lim\limits_{\omega_o\rightarrow\infty} \rho_1(\omega,\omega_o) \cong 
-\frac{1}{2} \left( \log{ \frac{\mathcal{Z}_1}{\mathcal{Z}_2} } \right) 
\frac{ \int_0^{\ell} g(x) e^{-2\pi i \left( \omega \left/ \omega_c \right. \right) x \left/ \ell \right. } \, dx }{ \int_0^{\ell} g(x) \, dx } ,
\end{equation}
where, from Eq.~(\ref{eq:Z0}), we note $\mathcal{Z}(0,\omega_o)\rightarrow\mathcal{Z}_1$ as $\omega_o\rightarrow\infty$. This all but eliminates $\omega_c$ from the expression of the reflection response, except for its appearance in the Fourier-like integral of the numerator, where it acts to delineate the region of high reflections, $\rho^{(HP)}_1(0)=(1/2)\log{ \left( \mathcal{Z}_2 / \mathcal{Z}_1 \right) }$, from that of low reflections, $\rho^{(HP)}_1(\infty)=0$. An estimate of the passband is to describe it as the range of frequencies $\omega$ such that $\omega_c/2\pi<\omega<\infty$. From Eq.~(\ref{eq:Z}), the corresponding characteristic impedance is
\begin{equation} \label{eq:Z-hp}
\mathcal{Z}^{(HP)}(x) = 
\lim\limits_{\omega_o\rightarrow\infty} \mathcal{Z}(x,\omega_o) = 
\mathcal{Z}_2 \exp{ \left[ \left( \log{ \frac{ \mathcal{Z}_1 }{ \mathcal{Z}_2 } } \right) 
\frac{ \int_x^{\ell} g(x') dx' }{ \int_0^{\ell} g(x') dx' } \right] } .
\end{equation}
The form of Eqs.~(\ref{eq:rho1-hp}) and (\ref{eq:Z-hp}) indicates that $g(x)$ of the lossless high-pass transformer may also be expressed as $g(x)=\alpha \; d \log{\mathcal{Z}^{(HP)}(x)} / dx$, where $\alpha$ is a constant. From Eq.~(\ref{eq:g}), we have an alternative expression of the boundary conditions of $g(x)$, viz.
\begin{subequations} \label{eq:g-alternative}
\begin{eqnarray}
{ \left. \frac{d}{dx} \log{\mathcal{Z}^{(HP)}(x)} \right| }_{x=0} = 0 , \;\;
{ \left. \frac{d}{dx} \log{\mathcal{Z}^{(HP)}(x)} \right| }_{x=\ell} = 0 , \label{eq:g-alternative-1} \\
{ \left. \frac{d^2}{dx^2} \log{\mathcal{Z}^{(HP)}(x)} \right| }_{x=0} = \frac{1}{\alpha} , \;\; 
{ \left. \frac{d^2}{dx^2} \log{\mathcal{Z}^{(HP)}(x)} \right| }_{x=\ell} = 0 . \label{eq:g-alternative-2}
\end{eqnarray}
\end{subequations}

\begin{figure}
\includegraphics[width=240pt, height=294pt]{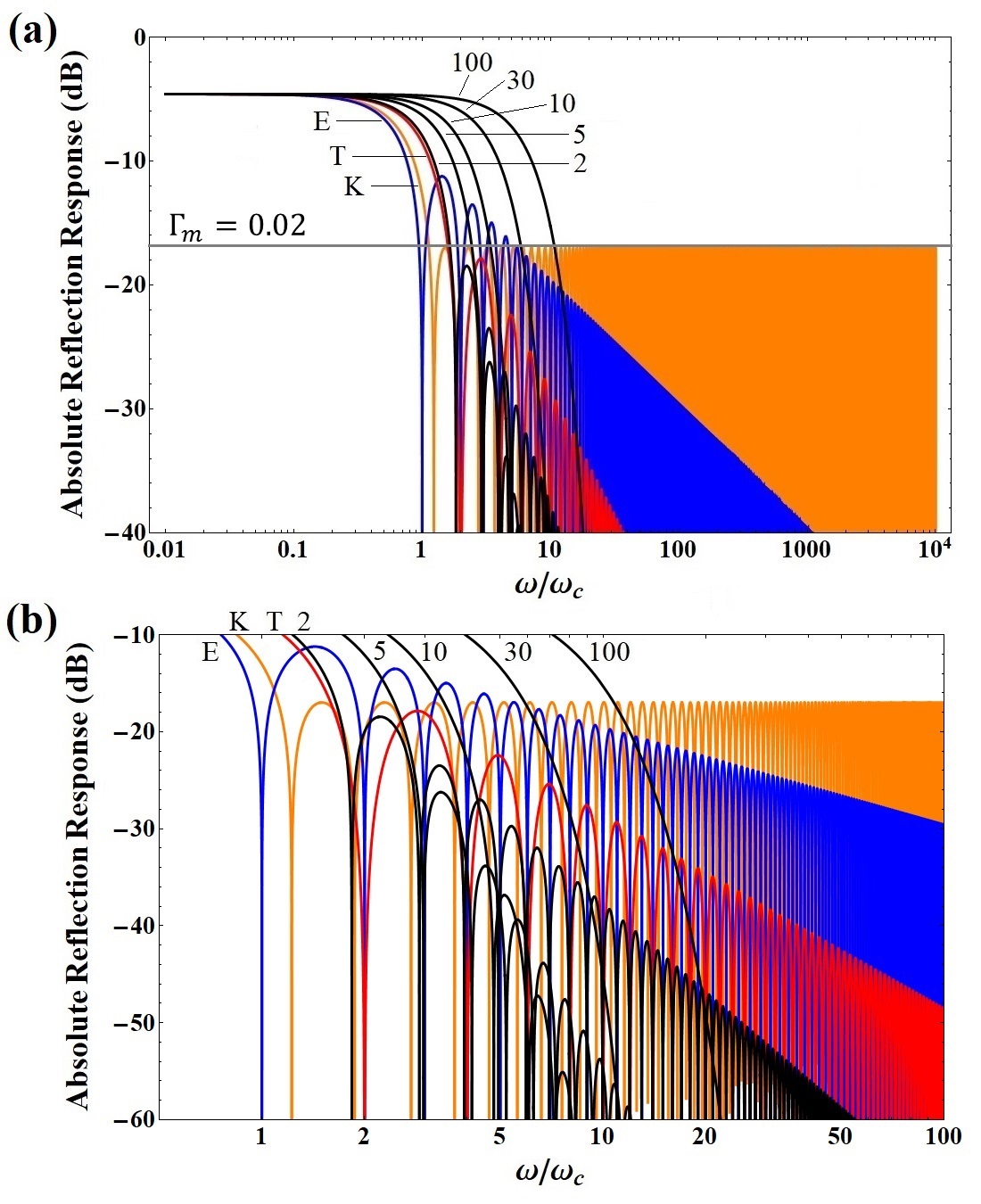}
\caption{\label{fig2} Input-line absolute reflection response (in decibels) versus $\omega/\omega_c$ (on logarithmic scale) for different wide-band high-pass lossless transformers: exponential (E), triangular (T), and Klopfenstein (K) tapers of Pozar,\cite{Pozar2012} as well as several $2N$-degree-polynomial tapers (numeric labels) discussed in text. (a) Depicts absolute reflection response across spectrum, with bandpass corresponding to $\omega>\omega_c$ and Klopfenstein parameter $\Gamma_m=0.02$. (b) Shows absolute reflection response at increased scale. In all cases $\mathcal{Z}_1=50$~$\Omega$, $\mathcal{Z}_2=100$~$\Omega$, $\ell=50$~mm, and $\omega_c/2\pi=100$~MHz.}
\end{figure}

As mentioned, the Fourier-transform-like integral in the numerator of Eq.~(\ref{eq:rho1-hp}) defines the high-pass frequency regime to be $\omega_c/2\pi<\omega<\infty$. The choice of $g(x)$ determines the extent to which $\left| \rho^{(HP)}_1(\omega) \right|$ is negligible over this interval. A good choice for $g(x)$ may be obtained by examining the asymptotic expansion of the integral for $2\pi\omega\gg\omega_c$. After repeated integration by parts $N$ times this expansion may be expressed as
\begin{multline} \label{eq:expansion-hp}
\int_0^{\ell} g(x) e^{-2\pi i \left( \omega \left/ \omega_c \right. \right) x \left/ \ell \right. } \, dx =
{ \left( \frac{\omega_c}{ 2\pi \omega} \right) }^2 + 
\sum^N_{n=3} { \left( \frac{ i \omega_c }{ 2\pi \omega} \right) }^n 
\left[ g^{(n-1)}(\ell) e^{-2\pi i \left( \omega \left/ \omega_c \right. \right) } - g^{(n-1)}(0) \right]  \\
+ { \left( \frac{ i \omega_c }{ 2\pi \omega} \right) }^N \int_0^\ell g^{(N)}(x) 
e^{-2\pi i \left( \omega \left/ \omega_c \right. \right) x \left/ \ell \right. } \, dx ,
\end{multline}
where $g^{(n)}(x)$ refers to the n-th derivative with respect to $x$ of $g(x)$, and the integral on the right side of the equation is the expansion remainder. A good high-pass transformer is one with a $g(x)$ that eliminates the second-order term in $\omega_c/(2\pi\omega)$ on the right side of Eq.~(\ref{eq:expansion-hp}); a better one also eliminates the third-order term, and so on.

In Appendix~\ref{appendix:HighPass} we demonstrate how a $2N$-degree polynomial choice for $g(x)$ can be used to eliminate terms of Eq.~(\ref{eq:expansion-hp}) to order $N$ in $\omega_c/(2\pi\omega)$. Using this $2N$-degree polynomial as our $g(x)$, we obtain a reflection response $\rho^{(HP)}_1(\omega,N)$ and characteristic impedance $\mathcal{Z}^{(HP)}(x,N)$ expressable as
\begin{equation} \label{eq:rho1-N-hp-1}
\rho^{(HP)}_1(\omega,N) \cong 
-\frac{ \Gamma\left( 3/2+N \right) }{\sqrt{\pi}} \left( \log{ \frac{\mathcal{Z}_1}{\mathcal{Z}_2} } \right) 
{ \left( \frac{2\omega_c}{\pi\omega} \right) }^N 
j_N \left( \pi\omega/\omega_c \right) \, e^{-\pi i \left( \omega / \omega_c \right)} ,
\end{equation}
\begin{equation} \label{eq:Z-N-hp-1}
\mathcal{Z}^{(HP)}(x,N) = \mathcal{Z}_2 
\exp{ \left[ \left( \log{\frac{ \mathcal{Z}_1 }{ \mathcal{Z}_2 }} \right) I\left( 1 - x/\ell;N+1,N+1 \right) \right] } ,
\end{equation}
respectively, where $\Gamma(z)$ is the gamma function, $j_N(z)$ is a spherical Bessel function, and
\begin{equation} \label{beta-function}
I\left(z;N+1,N+1 \right) = \frac{ \left( 2N + 1 \right)! }{ { \left( N! \right) }^2 } \int_0^z { \left( u - u^2 \right) }^N du 
\end{equation}
is a regularized incomplete beta function.\cite[p.~263]{AbramowitzAndStegun} By construction we have $|\rho^{(HP)}_1(\omega,N)|\propto 1/\omega^{N+1}$ as $\omega\rightarrow\infty$; conversely, as $\omega\rightarrow 0$, we find $|\rho^{(HP)}_1(0,N)|\cong (1/2)|\log{ \left( \mathcal{Z}_1/\mathcal{Z}_2 \right) }|$.

We illustrate our results with the specific example of a wide-band high-pass lossless transformer placed between a $50$~$\Omega$ input line and $100$~$\Omega$ load-bearing line. In this example our goal is a design with bandpass of $1$-$10$~GHz, appropriate for a supconducting parametric amplifier, with high-frequency asymptotic reflections damped as strongly as possible. For concreteness, assume a transformer length of $\ell=50$~mm, with $L_1=2.5$~pH/$\mu$m, $C_1=0.001$~pF/$\mu$m, $L_2=10$~pH/$\mu$m, and $C_2=0.001$~pF/$\mu$m. In this case the characteristic frequency of the transformer is $\omega_c/2\pi=0.5/(\ell\sqrt{L_2 C_2})=100$ MHz, but this frequency can be adjusted by changing the value of $\ell$.

In Fig.~\ref{fig2}(a) we plot the absolute reflection response $|\rho^{(HP)}_1(\omega,N)|$ obtained from Eq.~(\ref{eq:rho1-N-hp-1}) as a function of the ratio of the input-signal frequency $\omega$ to the transformer characteristic frequency $\omega_c$, on a logarithmic scale, for design values $N=2$, $5$, $10$, $30$, and $100$. For comparison we also plot the absolute reflection response of the exponential, triangular, and Klopfenstein tapers that are described in Pozar.\cite{Pozar2012} Our example is constructed to match the example in the Pozar text as closely as possible. In particular, for the Klopfenstein-taper parameters, corresponding to a ripple of 2\%, we set $\Gamma_0=0.346574$, $\Gamma_m=0.02$, and $A=3.54468$. For the Klopfenstein characteristic impedance we used the formula given by Pozar,\cite{Pozar2012} viz.
\begin{equation} \label{eq:klopfenstein-Z}
\mathcal{Z}^{(K)}(x) = \sqrt{\mathcal{Z}_2 \, \mathcal{Z}_2} 
\exp{ \left[ \Gamma_m A \int_0^{2x/\ell-1} \frac{I_1\left( A \sqrt{1-y^2} \right)}{\sqrt{1-y^2}} dy \right] } ,
\end{equation}
where $I_1(z)$ is a modified Bessel function of integer order. For the reflection response, we applied Eq~(\ref{eq:klopfenstein-Z}) to Eq.~(\ref{eq:diffeq}) and solved for the reflection coefficient, then set $x=0$ to obtain the input-line reflection response. The result may be expressed as
\begin{equation} \label{eq:klopfenstein-rho1}
\rho^{(K)}_1(\omega) = \frac{1}{2} \int_0^\ell \left[ \frac{d}{dx'} \log{ \mathcal{Z}^{(K)}(x') } \right] 
e^{-2\pi i \left( \omega \left/ \omega_c \right. \right) x' \left/ \ell \right. } dx'
= \Gamma_m \, e^{-i\pi \omega \left/ \omega_c \right. } \cos{ \left[ \sqrt{ { \left( \frac{\pi\omega}{\omega_c} \right) }^2 - A^2} \right] } ,
\end{equation}
where the last step follows from the Klopfenstein ansatz.\cite{Klopfenstein1956} Note that the Klopfenstein model is not optimal because the characteristic impedance $\mathcal{Z}^{(K)}(x)$ of Eq.~(\ref{eq:klopfenstein-Z}) does not satisfy all of the boundary conditions of Eq.~(\ref{eq:g-alternative}). Similarly, one can show that both the triangular and exponential models are not optimal because their respective characteristic impedances also do not satisfy Eq.~(\ref{eq:g-alternative}).

In Fig.~\ref{fig2}(b), we show the drop-off of absolute reflection response for a range of input frequencies starting from $\omega=\omega_c$. This provides a comparison of the ripple response of the different taper designs. The $2N$-degree polynomial tapers have the smallest residual ripple, even for the case of $N=2$, owing to the built-in $1/\omega^{N+1}$ behavior as $\omega\rightarrow\infty$. At $\omega/\omega_c=100$, the maximum absolute reflection responses of the exponential and triangular tapers are approximately $-29$~dB and $-48$~dB, respectively, while the Klopfenstein ripple retains a fixed maximum of 2\%, as per the $\Gamma_m=0.02$ parameter setting, i.e., $-17$~dB. At $\omega/\omega_c=100$ the $2N$-degree polynomial cases of $N=2$, $N=5$, $N=10$, $N=30$, and $N=100$ are all well below $-60$~dB. 

For the superconducting parametric amplifiers discussed in the introduction, in the operating range of $1$-$10$~GHz, the highly-damped reflections of the $2N$-degree polynomial design can substantially limit signal-gain ripple, as induced by impedance mismatch. A caveat of the $2N$-degree polynomial design is that as $N$ increases the lower bound of the passband tends to shift to higher frequencies, as is evident in Fig.~\ref{fig2}(a). In particular, from Eq.~(\ref{eq:rho1-poly-lessthan}) of Appendix~\ref{appendix:HighPass}, we have
\begin{equation}  \label{eq:rho1-poly-lessthan-1}
\lim_{N\rightarrow\infty} \left| \rho^{(HP)}_1(\omega,N) \right| \cong \frac{1}{2} \left| \log{ \frac{\mathcal{Z}_1}{\mathcal{Z}_2} } \right| ,
\end{equation}
indicative of the passband diminishing to zero width as $N\rightarrow\infty$. This result is consistent with the Bode-Fano criterion\cite{Bode1945,*Fano1950-1} in the sense that as $N\rightarrow\infty$  one might suspect the passband becoming a region of perfectly zero reflections since $|\rho^{(HP)}_1(\omega,N)|\propto 1/\omega^{N+1}$ as $\omega\rightarrow\infty$; however, the order in which one takes limits matters, so as $N\rightarrow\infty$ for arbitrary $\omega$ the bandwidth instead goes to zero. For modest increases in $N$, the tendency for the lower bound of the passband to shift to higher frequencies can be compensated for in the taper design by increasing the length $\ell$ of the taper, thereby decreasing $\omega_c$.

\begin{figure}
\includegraphics[width=240pt, height=156pt]{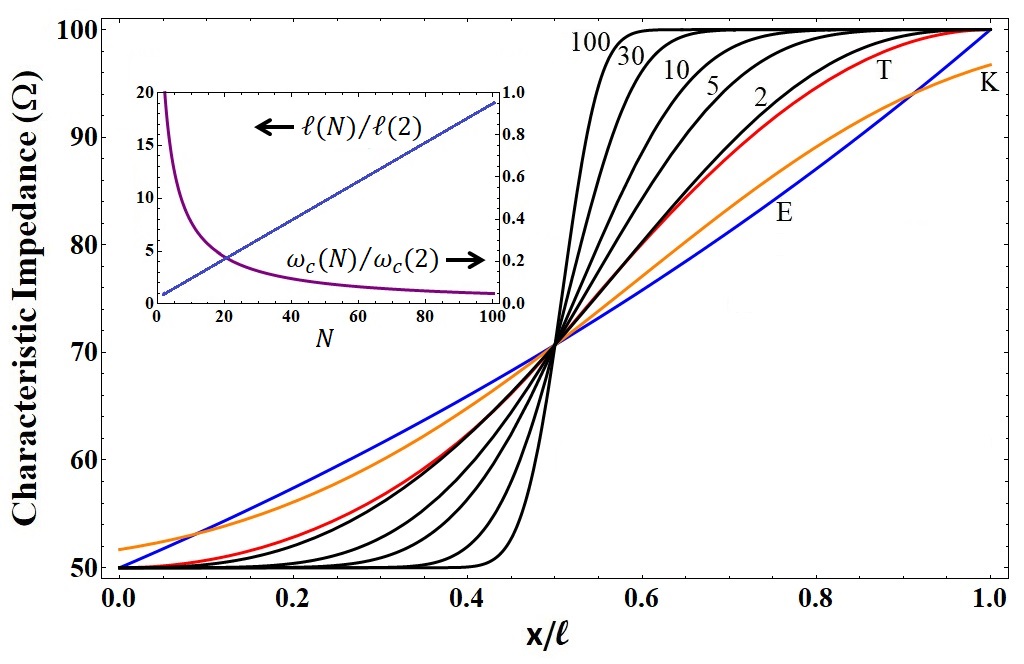}
\caption{\label{fig3} Characteristic impedance as function of $x/\ell$ along taper for wide-band high-pass lossless transformers: exponential (E), triangular (T), and Klopfenstein (K), as well as several $2N$-degree-polynomial tapers (numeric labels). As discussed in text, to compare a $2N$-degree polynomial taper ($N\ge 2$) to other taper designs, the lowest frequency of the $2N$-degree-polynomial passband, $\omega_1(N)$ of Eq.~(\ref{eq:hp-lower-bound}), is held fixed, such that $\omega_1(N)=\omega_1(2)$ for all $N>2$. The inset is a plot of Eq.~(\ref{eq:hp-constraints}), showing taper length $\ell(N)$ (left vertical axis), or equivalently characteristic frequency $\omega_c(N)$ (right vertical axis), as a function of $N$.}
\end{figure}

In Fig.~\ref{fig3}, we plot the characteristic impedance, corresponding to the the absolute reflection response of Fig~\ref{fig2}, as a function of relative position $x/\ell$ within the taper. The characteristic impedance $\mathcal{Z}^{(HP)}(x,N)$ of Eq.~(\ref{eq:Z-N-hp-1}) is shown for design values of $N=2$, $5$, $10$, $30$, and $100$, while that of the exponential, triangular, and Klopfenstein tapers is as in the Pozar text.\cite{Pozar2012} Recall from Fig.~\ref{fig2}(a) and the discussion of Eq.~(\ref{eq:rho1-poly-lessthan-1}) that the lower bound of the passband shifts to higher frequencies as $N$ increases. To maintain a fixed lower bound we must allow $\ell$ to increase accordingly, i.e., $\omega_c=\pi v_g/\ell$ becomes smaller. Specifically, to compare the characteristic impedance of a $2N$-degree polynomial taper ($N\ge 2$) to that of the other taper designs in Fig.~\ref{fig3}, we fix the lowest frequency of the $2N$-degree-polynomial passband, $\omega_1(N)$, such that $\omega_1(N)=\omega_1(2)$ for all $N>2$. In this way, all of the $2N$-degree polynomial tapers will have the same bandwidth as the $N=2$ case, comparable to the bandwidths of the exponential, triangular, and Klopfenstein tapers. Since the lower bound of the Klopfenstein passband is typically defined as the frequency corresponding to the first zero of reflections, we may define $\omega_1(N)$ similarly. Thus, from Eq.~(\ref{eq:rho1-N-hp-1}), the first zero of $|\rho^{(HP)}_1(\omega,N)|$ is the first zero of the spherical Bessel function $j_N(\pi\omega/\omega_c)$, call it $z_N$, i.e., $j_N(z_N)\equiv 0$ such that
\begin{equation} \label{eq:hp-lower-bound}
\omega_1(N) = \frac{z_N}{\pi} \, \omega_c(N) ,
\end{equation}
where $\omega_c(N)=\pi v_g / \ell(N)$. For example, first zeros of the spherical Bessel function include $z_2=5.76346$, $z_5=9.35581$, and $z_{10}=15.0335$. Then, setting $\omega_1(N)=\omega_1(2)$ implies
\begin{equation} \label{eq:hp-constraints}
\omega_c(N) = \frac{z_2}{z_N} \, \omega_c(2) , \,\, \ell(N) = \frac{z_N}{z_2} \, \ell(2) .
\end{equation}
The inset of Fig.~\ref{fig3} is a plot of Eq.~(\ref{eq:hp-constraints}) as a function of increasing $N$, showing both $\ell(N)$, corresponding to the left vertical axis, and $\omega_c(N)$, corresponding to the right vertical axis. For example, in order that the $N=100$ case have the same passband as the $N=2$ case, $\ell(100)$ must be approximately $20$ times greater than $\ell(2)$. In particular, as $N\rightarrow\infty$ then $\ell(N)\rightarrow\infty$, corresponding to a taper of unattainably infinite length. Moreover, all of the curves corresponding to $N\gg 1$ are difficult to realize physically given the length of transformer required, as well as the material composition and other geometric factors, i.e., $v_g$, that contribute to $\omega_c(N)$.

In Fig.~\ref{fig3}, note that the characteristic impedance of the $N=2$ case of the $2N$-polynomial taper design closely matches that of the triangular taper, but with a more strongly damped reflection oscillation, due to the $1/\omega^3$ behavior of the input-line reflections as $\omega\rightarrow\infty$. As $N$ increases the damping of these reflections increases as $1/\omega^{N+1}$, and the shape of the characteristic impedance approaches a sharper profile for $x\approx \ell/2$. The point $\mathcal{Z}^{(HP)}(\ell/2,N)=\sqrt{\mathcal{Z}_1 \mathcal{Z}_2}\cong 70.71$~$\Omega$ is fixed for all $N$. 

In Fig.~(\ref{fig3}), the sharpness of the characteristic impedance profile as $N\rightarrow\infty$ makes it difficult to fabricate such a taper due to limitations of line composition and geometry, as mentioned earlier. Even for the case of the $N=100$ taper, where from Fig.~\ref{fig2}(a) the reflection response is negligible for $\omega/\omega_c>100$, there is a sharp change in the characteristic impedance for $x\approx \ell/2$ that would make this taper extremely challenging to fabricate, given the necessary length of the taper. A compromise is to select a value of $N$ between the $N=2$ and $N=100$ cases. Clearly, a good choice is $N=2$, but a better choice might be $N=5$ or $N=10$. The best choice is to select the largest $N$ permitted by the fabrication constraints of the application such that residual oscillations are damped to the greatest extent possible within the region of exploitable passband---this is the physical limit of the high-pass design for the application.

Important points regarding the optimal wide-band high-pass transformer are:
\begin{enumerate}
\item By setting the design frequency $\omega_o$ of the transformer to infinity we realized a high-pass bandwidth the order of $\omega_c/2\pi<\omega<\infty$.
\item To exploit this bandwidth to its greatest extent we expanded the Fourier-like integral of the numerator of Eq.~(\ref{eq:rho1-hp}) in an asymptotic series, and we choose $g(x)$ so as to eliminate the lowest $N-1$ terms of this series, as described in detail in Appendix~\ref{appendix:HighPass}; this defined a model parametrized by integer $N$, where the reflection response, by construction, is $|\rho^{(HP)}_1(\omega,N)|\propto 1/\omega^{N+1}$ as $\omega\rightarrow\infty$.
\item The optimal characteristic impedance of the optimal high-pass-transformer model, $\mathcal{Z}^{(HP)}(x,N)$, is given by Eq.~(\ref{eq:Z-N-hp-1}), as derived in Appendix~\ref{appendix:HighPass}.
\item We compared this optimal high-pass transformer model to the Klopfenstein, triangular, and exponential models, demonstrating its superior reflection response at high frequencies; this resulted from the optimal form of Eq.~(\ref{eq:rho1-hp}), which allowed us to design the asymptotic behavior.
\item The Klopfenstein, triangular, and exponential models are not optimal transformers because their respective characteristic impedances do not satisfy the boundary conditions of Eq.~(\ref{eq:g-alternative}).
\item A limitation on the size of integer $N$ of the optimal design is imposed by the physical composition and geometry of the taper; a value in the range of of $2\le N \le 10$ is probably reasonable for most applications.
\end{enumerate}

\subsection{\label{subsection:lowpass}The Wide-Band Low-Pass Impedance Transformer}
In a manner analogous to the high-pass transformer, recall from our introductory remarks that a wide-band low-pass impedance transformer can be constructed if the design frequency $\omega_o$ is detuned from $\omega_c$ such that $\omega_o\rightarrow 0$. In this case Eq.~(\ref{eq:rho1-w}) becomes
\begin{equation} \label{eq:rho1-lp}
\rho^{(LP)}_1(\omega) = 
\lim\limits_{\omega_o\rightarrow 0} \rho_1(\omega,\omega_o) \cong 
\frac{1}{2 \ell^2} \left[ \log{ \frac{\mathcal{Z}(0,0)}{\mathcal{Z}_2} } \right] { \left( \frac{ 2\pi i \omega }{ \omega_c } \right) }^2 
\int_0^{\ell} g(x) e^{-2\pi i \left( \omega \left/ \omega_c \right. \right) x \left/ \ell \right. } dx ,
\end{equation}
where $\mathcal{Z}(0,0)$ is determined from Eq.~(\ref{eq:Z0}), for the case $\omega_o=0$, viz. 
\begin{equation} \label{eq:Z0-lp}
2 \left[ \mathcal{Z}_1 - \mathcal{Z}(0,0) \right] + \left[ \mathcal{Z}_1 + \mathcal{Z}(0,0) \right] \log{ \frac{\mathcal{Z}(0,0)}{\mathcal{Z}_2} } = 0 .
\end{equation}
As in Eq.~(\ref{eq:rho1-hp}) of the high-pass case, $\omega_c$ delineates the low-frequency and high-frequency regimes. From Eq.~(\ref{eq:Z}), the corresponding characteristic impedance is
\begin{equation} \label{eq:Z-lp}
\mathcal{Z}^{(LP)}(x) = \lim\limits_{\omega_o\rightarrow 0} \mathcal{Z}(x,\omega_o) = 
\mathcal{Z}_2 \exp{ \left\{ \left[ \log{ \frac{ \mathcal{Z}(0,0) }{ \mathcal{Z}_2 } } \right] \left[ \frac{ d g(x) }{ dx } \right] \right\} } .
\end{equation}

\begin{figure}
\includegraphics[width=240pt, height=301pt]{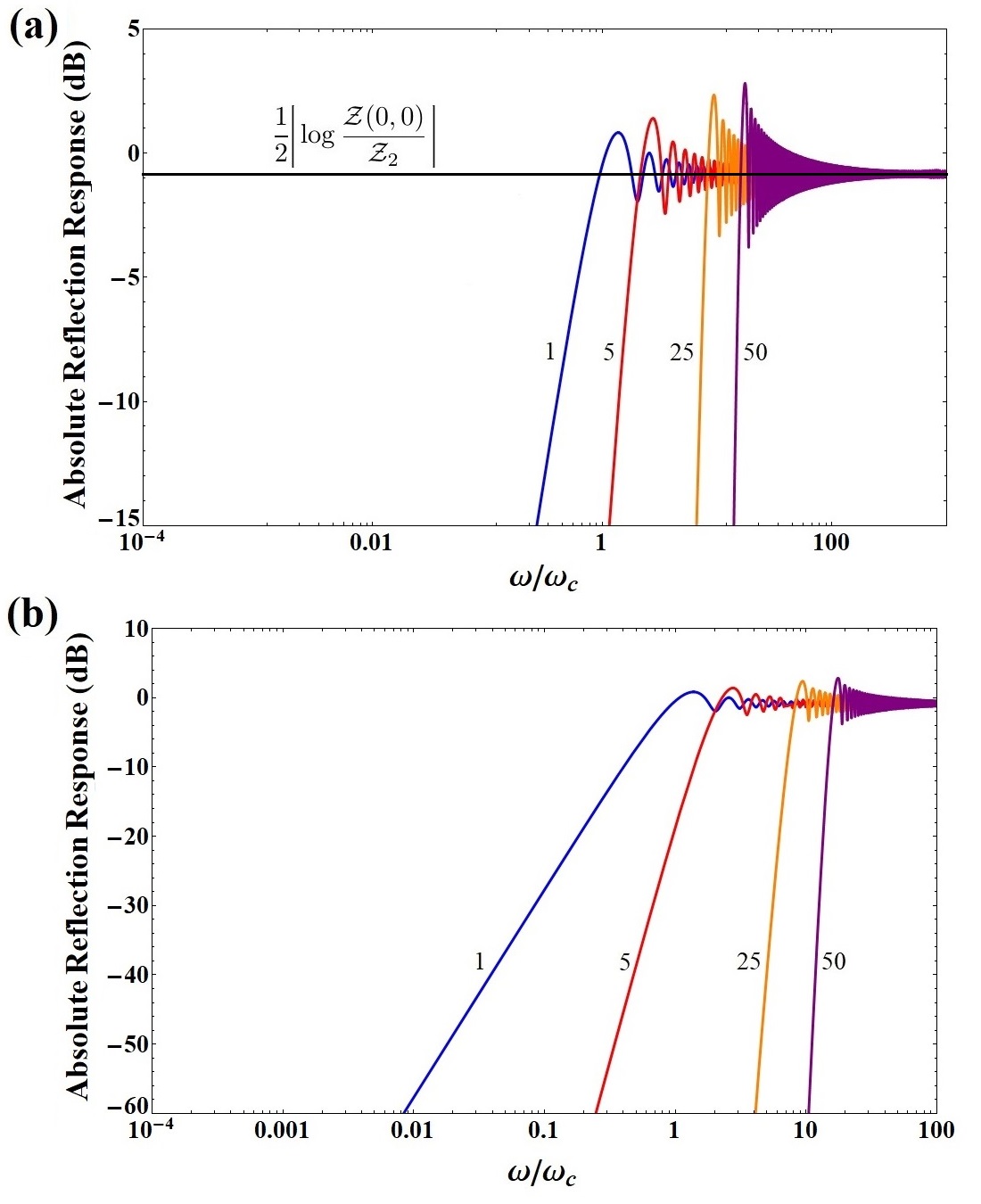}
\caption{\label{fig4} Input-line absolute reflection response (in decibels) versus $\omega/\omega_c$ (on logarithmic scale) for wide-band low-pass lossless transformers of $(N+3)$-degree-polynomial taper design discussed in text. Horizontal line represents high-frequency asymptote $(1/2)|\log{\left[ \mathcal{Z}(0,0) / \mathcal{Z}_2 \right]}|$ as $\omega\rightarrow\infty$. (a) Shows absolute reflection response across frequency spectrum, with bandpass corresponding to $\omega<\omega_c$. (b) Depicts absolute reflection response at increased scale, indicating smallness of response over range of frequencies of interest. In all cases $\mathcal{Z}_1=50$~$\Omega$, $\mathcal{Z}_2=100$~$\Omega$, $\ell=5$~mm, and $\omega_c/2\pi=1$~GHz.}
\end{figure}

As in the high-pass case, we can determine a choice for $g(x)$ by considering the asymptotic behavior of the Fourier-transform-like integral of Eq.~(\ref{eq:rho1-lp}) as $\omega\rightarrow 0$. Assuming $2\pi\omega\ll\omega_c$, and repeatedly integrating by parts $N$ times, we obtain
\begin{multline} \label{eq:expansion-lp}
\int_0^{\ell} g(x) e^{-2\pi i \left( \omega \left/ \omega_c \right. \right) x \left/ \ell \right. } dx = 
e^{-2\pi i \left( \omega \left/ \omega_c \right. \right) } \sum\limits_{n=0}^{N-1} { \left( \frac{2\pi i\omega}{\omega_c} \right) }^n \int_0^{\ell} G^{(n)}(x) dx \\
+ { \left( \frac{2\pi i\omega}{\omega_c} \right) }^{N-1} \int_0^{\ell} G^{(N)}(x) e^{-2\pi i \left( \omega \left/ \omega_c \right. \right) x/\ell }  dx ,
\end{multline}
where the last term is the expansion remainder, and we have defined
\begin{equation}
G^{(n)}(x) = \frac{1}{\ell} \int_0^x G^{(n-1)}(x') \, dx' \; ; \;\; G^{(0)}(x) = g(x) .
\end{equation}

In Appendix \ref{appendix:LowPass} we demonstrate how a $(N+3)$-degree polynomial choice for $g(x)$ can be used to eliminate terms of Eq.~(\ref{eq:expansion-lp}) to order $N-1$ in $2\pi\omega/\omega_c$. In this case the reflection response $\rho^{(LP)}_1(\omega,N)$ and characteristic impedance $\mathcal{Z}^{(LP)}(x,N)$ are
\begin{multline} \label{eq:rho1-N-lp}
\rho^{(LP)}_1(\omega,N) \cong
\frac{1}{2} \frac{ \left( N + 2 \right)! }{ \left( 2N + 4 \right)! } \left[ \log{ \frac{\mathcal{Z}(0,0)}{\mathcal{Z}_2} } \right] \\
\times { \left( \frac{ 2\pi i \omega }{ \omega_c } \right) }^{N+2} M \left( N + 3, 2N + 5 , \frac{ 2\pi i \omega }{ \omega_c } \right) 
e^{-2\pi i \left( \omega \left/ \omega_c \right. \right) } ,
\end{multline}
\begin{equation} \label{eq:Z-N-lp}
\mathcal{Z}^{(LP)}(x,N) = \mathcal{Z}_2 
\exp{ \left\{ \left[ \log{ \frac{ \mathcal{Z}(0,0) }{ \mathcal{Z}_2 } } \right] P^{(0,-1)}_{N+2}(1-2x/\ell) \right\} } ,
\end{equation}
respectively, where $M(a,b,z)$ is Kummer's confluent hypergeometric function\cite[p.~504]{AbramowitzAndStegun} and $P^{(\alpha,\beta)}_n(z)$ is a Jacobi polynomial of order $n$.\cite[p.~561]{AbramowitzAndStegun} By construction $|\rho^{(LP)}_1(\omega,N)|\propto \omega^{N+2}$ as $\omega\rightarrow 0$; conversely, as $\omega\rightarrow\infty$, we find $|\rho^{(LP)}_1(\omega,N)|\rightarrow (1/2)|\log{ \left[ \mathcal{Z}(0,0)/\mathcal{Z}_2 \right] }|$.

As in the high-pass case, we illustrate the wide-band low-pass lossless transformer with the specific example of a $50$~$\Omega$ input line and $100$~$\Omega$ load-bearing line. For concreteness, assume a transformer length of $\ell=5$~mm, with $L_1=2.5$~pH/$\mu$m, $C_1=0.001$~pF/$\mu$m, $L_2=10$~pH/$\mu$m, and $C_2=0.001$~pF/$\mu$m. Therefore, we have a bandpass that extends up to the transformer characteristic frequency, $\omega_c/2\pi=0.5/(\ell\sqrt{L_2 C_2})=1$ GHz. Again, this frequency can be adjusted by changing the length $\ell$ of the transformer.

\begin{figure}
\includegraphics[width=240pt, height=156pt]{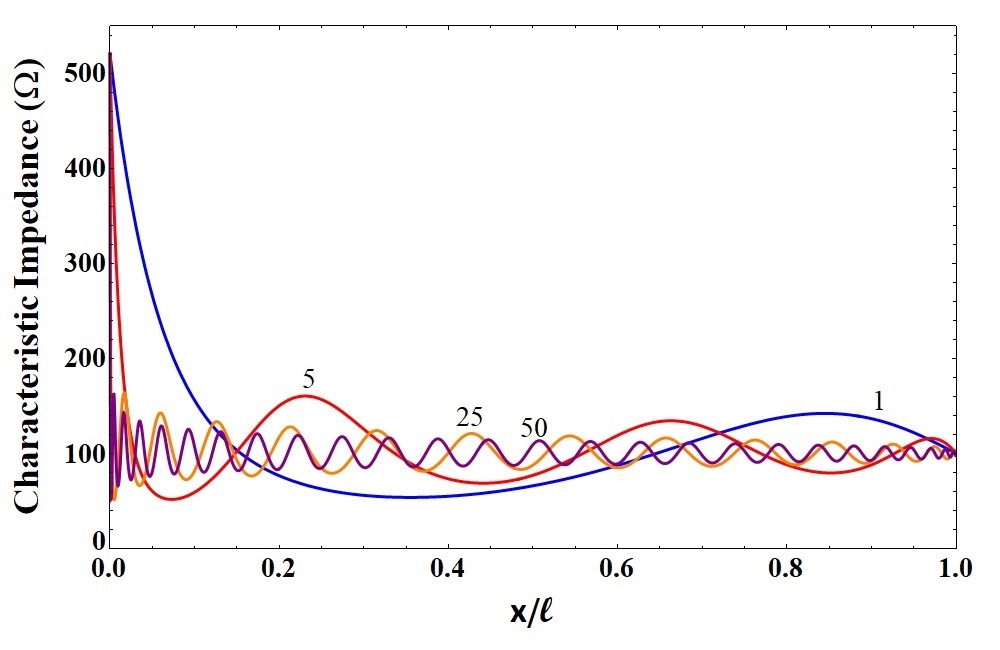}
\caption{\label{fig5} Characteristic impedance of the wide-band low-pass lossless transformers of $(N+3)$-degree polynomial design discussed in text, plotted as a function of position $x$ along length $\ell$ of transformer. Here, $\mathcal{Z}_1=50$~$\Omega$ and $\mathcal{Z}_2=100$~$\Omega$.}
\end{figure}

In Fig.~\ref{fig4}(a) we plot the absolute reflection response $|\rho^{(LP)}_1(\omega,N)|$, as obtained from Eq. (\ref{eq:rho1-N-lp}), as a function of the ratio of the input-signal frequency $\omega$ to the transformer characteristic frequency $\omega_c$, on a logarithmic scale, for design values of $N=1$, $5$, $25$, and $50$. The figure shows negligible reflection response over a bandpass of up to $\omega\cong\omega_c$, followed by a region $\omega\approx\omega_c$ characterized by large oscillating reflections. For $\omega\approx\omega_c$ the magnitude of reflections surpasses unity by as much as a factor of two due to the breakdown of the small-reflection approximation of Eqs.~(\ref{eq:diffeq}) and (\ref{eq:rho1-w}). This could be addressed by solving the full Riccati equation for the voltage reflection coefficient. However, the small-reflection approximation of Eq. (\ref{eq:diffeq}) is more than adequate to estimate the breadth of the transformer pass band. In particular, we see from Fig.~\ref{fig4}(b) the smallness of the bandpass reflections as the frequency decreases from $\omega\approx\omega_c$.

In Fig.~\ref{fig5} we plot the characteristic impedance of the wide-bandwidth low-pass lossless transformer as a function of relative position $x/\ell$ within the taper, for values of $N=1$, $5$, $25$, and $50$. The figure illustrates the discontinuity in characteristic impedance at the $x=0$ interface that is particular to the low-pass transformer, independent of taper design choice. In this example the input line characteristic impedance is $50$~$\Omega$ and just inside the taper we have $\mathcal{Z}\left(0,0\right)\approx 500$~$\Omega$. The discontinuity is governed by the size of the impedance mismatch between input and load-bearing lines, with solution for $\mathcal{Z}(0,0)$ obtained from Eq.~(\ref{eq:Z0-lp}). As the impedance mismatch increases, fabrication of the transformer becomes a challenge because the taper characteristic impedance sharply increases at $x=0$. This is the principal difficulty in building the low-pass impedance transformer, regardless of design choice.

Nevertheless, the solutions of the $(N+3)$-degree polynomial design indicate the direction to take in fabricating the low-pass transformer, if the impedance mismatch is not too great. In Fig.~\ref{fig5}, as $N$ increases the $(N+3)$-degree polynomial solution incurs greater undulation, which again presents fabrication difficulties. From Eq.~(\ref{eq:darboux}), the undulation behavior of $N\gg 1$ may be approximated as
\begin{equation}
\mathcal{Z}^{(LP)}(x) \cong \mathcal{Z}_2 
\exp{ \left\{ \left[ \log{ \frac{ \mathcal{Z}(0,0) }{ \mathcal{Z}_2 } } \right] 
\frac{\cos{ \left[ N \varphi(x) - \pi/4 \right] }}{ \sqrt{ \pi N \tan{ \left[ \varphi(x)/2 \right] } } } \right\} } ,
\end{equation}
where $\varphi(x) = \arccos{ \left( 1-2x/\ell \right)}$. However, as Fig.~\ref{fig4} shows, the simplest case of $N=1$ has negligible reflection response over frequency band $0<\omega/\omega_c<0.1$, with corresponding characteristic impedance of Fig.~\ref{fig5} exhibiting a very smooth and gradual undulation---only representing a challenge to fabrication at the $x=0$ end of the taper. Therefore, the $N=1$ case can represent a feasible design for an optimal low-pass transformer. Similar to the polynomial designs of the high-pass transformer, the largest value of $N$ that can be accommodated by fabrication constraints represents the best physical taper design.

An interesting aspect to the low-pass transformer is its ability to act as a filter of high frequencies. In Fig.~\ref{fig4}(a), the black horizontal line is the asymptote for the limit of the absolute reflection response $|\rho^{(LP)}_1(\omega,N)|\rightarrow (1/2)|\log{ \left[ \mathcal{Z}(0,0)/\mathcal{Z}_2 \right]}|$, as input frequency $\omega\rightarrow\infty$, independent of $N$. When the asymptote approaches unity the low-pass transformer acts as low-pass filter, suppressing transmission of frequencies $\omega>\omega_c$. In this limit we can approximate $\mathcal{Z}(0,0)$ from Eq.~(\ref{eq:Z0-lp}) as $\mathcal{Z}(0,0)\cong\mathcal{Z}_2\exp{\left(2\right)}-4 \mathcal{Z}_1$, where $4\mathcal{Z}_1\ll\mathcal{Z}_2\exp{\left(2\right)}$. The absolute reflection response is then $|\rho^{(LP)}_1(\omega,N)|\cong |1 - 2 e^{-2} (\mathcal{Z}_1/\mathcal{Z}_2)|$ as $\omega\rightarrow\infty$. Thus, as the impedance mismatch between input line and load-bearing line increases, the filter becomes more effective. Efficacy of the device is limited by fabrication constraints imposed by the mismatch at the $x=0$ interface, as discussed earlier, but the device may have application to reduction of the Purcell effect (at $\sim 7$~GHz) in superconducting transmon and Xmon qubit-readout measurements.\cite{Reed2010,*Jeffrey2014,*BronnNT2015,*Sete2015}

Important points regarding the optimal wide-band low-pass transformer are:
\begin{enumerate}
\item By setting the design frequency $\omega_o$ of the transformer to zero we realized a low-pass bandwidth the order of $0<\omega<\omega_c/2\pi$.
\item Analogous to the high-pass case, we exploited this bandwidth to its greatest extent by expanding the Fourier-like integral of the numerator of Eq.~(\ref{eq:rho1-lp}) in an asymptotic series, choosing $g(x)$ so as to eliminate the lowest $N$ terms of this series, as described in detail in Appendix~\ref{appendix:LowPass}; this again defined a model parametrized by integer $N$, where the reflection response, by construction, is $|\rho^{(LP)}_1(\omega,N)|\propto \omega^{N+2}$ as $\omega\rightarrow 0$.
\item The optimal characteristic impedance of the optimal low-pass-transformer model, $\mathcal{Z}^{(LP)}(x,N)$, is given by Eq.~(\ref{eq:Z-N-lp}), as derived in Appendix~\ref{appendix:LowPass}.
\item We compared optimal low-pass transformer designs of different values of integer $N$, plotting reflection response versus frequency in Fig.~\ref{fig4} and characteristic impedance versus position along the taper in Fig.~\ref{fig5}; the small-reflection approximation employed in Fig.~\ref{fig4} tends to break down at high frequencies and larger values of $N$.
\end{enumerate}

\section{\label{section:remarks}Concluding Remarks}
We presented a variational approach to determine the optimal form of the reflection coefficient of a tapered impedance transformer of length $\ell$, for a specific design frequency $\omega_o$. We used this result to construct the characteristic impedance and input-line reflection response of an optimal lossless transformer, defining the optimal transformer design in terms of $\omega_o$ and a real-valued function $g(x)$ satisfying the boundary conditions of Eq. (\ref{eq:g}). The input-line reflection response was shown to depend on a characteristic frequency $\omega_c=\pi v_g/\ell$, where $v_g$ is the constant transformer group velocity.

By construction the input-line reflection response, as a function of arbitrary frequency $\omega$, is zero specifically at $\omega=\omega_o$, indicative of narrow pass band. However, we showed that if $\omega_o$ is detuned far from $\omega_c$ then, for an extended range of frequencies $\omega$ about $\omega_o$, the magnitude of input-line reflections is negligible. Specifically, when we took the limit $\omega_o\rightarrow\infty$ we obtained a high-pass transformer design with pass band $\omega_c\lesssim\omega<\infty$, where the input-line reflection response and characteristic impedance are given by Eqs. (\ref{eq:rho1-hp}) and (\ref{eq:Z-hp}), respectively. Similarly, when we took the limit $\omega_o\rightarrow 0$ we obtained a low-pass transformer design with pass band $0<\omega\lesssim\omega_c$, where input-line reflection response and characteristic impedance are given by Eqs. (\ref{eq:rho1-lp}) and (\ref{eq:Z-lp}), respectively.

Having derived a general form for wide-bandwidth transformers in terms of $g(x)$, for both the high-pass and low-pass frequency regimes, we then showed, for each regime, how to choose a polynomial $g(x)$ that produces the widest exploitable pass band possible. In the case of the high-pass transformer, we compared our results to existing optimal taper designs, specifically the exponential, triangular, and Klopfenstein tapers described in Pozar.\cite{Pozar2012} We showed that are our design exhibits superior pass-band characteristics. For the case of the low-pass transformer, we demonstrated the inherent difficulty of fabrication of the $x=0$ end of the taper, due to the discontinuity of the characteristic impedance at this interface. Nevertheless, we proposed the $N=1$ case of our design as the simplest to fabricate, with greater efficacy the smaller the impedance mismatch of the application.

\begin{figure}
\includegraphics[width=240pt, height=128pt]{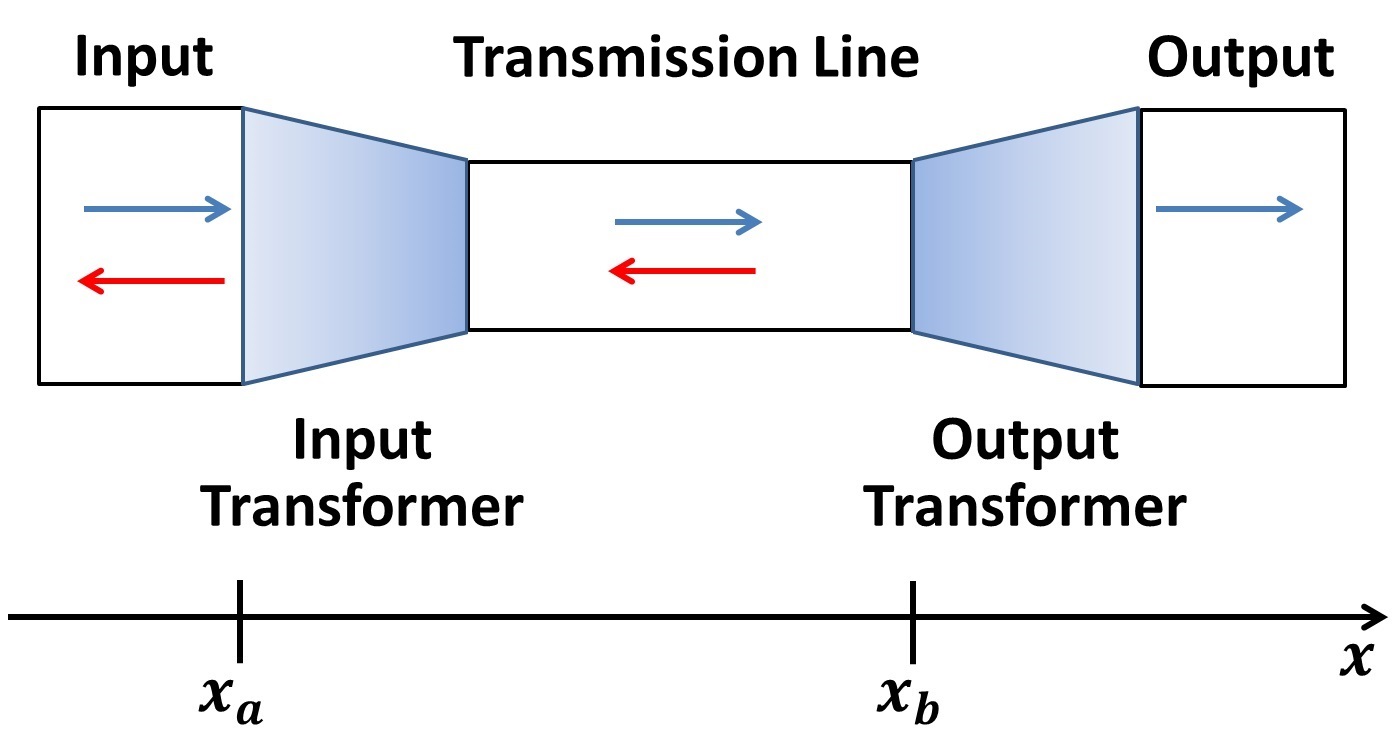}
\caption{\label{fig6} Schematic of transmission line with coupled transformers (shaded) at both input and output. Forward-traveling (reflected) signal is represented by arrow pointing to right (left). Forward-traveling signal of input line can reflect from input transformer at $x=x_a$. Inside transmission line, forward-traveling signal also can reflect from output transformer at $x=x_b$, thereby re-entering input transformer from right.}
\end{figure}

An important point to note is that our theory has focused on an isolated impedance transformer, one for which signals do not enter the transformer from the $x=\ell$ side. In practice, a transformer can be placed at both input and output of a load-bearing component, which means that the two transformers will be coupled, as in the example schematic of Fig.~\ref{fig6}. Just as a forward-traveling wave (arrow pointing to right) can reflect from the input transformer, upon transmission through the input transformer, a forward-traveling wave can also reflect from the output transformer, allowing this reflected signal to encounter the input transformer from the right, as pictured.

In the case the two coupled transformers of Fig.~\ref{fig6}, the optimization of their design requires simultaneous minimization of the voltage reflection coefficients at both $x=x_a$ and $x=x_b$, from the left. This can be performed in the manner of our variational approach, but also requires derivation of the corresponding coupled Riccati differential equations of the two tapers, as well as their boundary conditions. These equations can be obtained by extending the approach used in Appendix~\ref{appendix:DifferentialEquation}. We consider the case of coupled transformers to be an extension of our present work, and a subject of future focus.

As mentioned earlier, our motivation for the present study is to improve the signal-to-noise ratio of superconducting amplifiers used in quantum-information research. We are presently engaged in fabrication and validation of the transformer designs derived here, for the coplanar waveguides that comprise our amplifiers. One aim of our experimental analysis is to ascertain the limit of fabrication techniques to capture and leverage improvements implied by these designs, i.e., whether these transformers may be fabricated to sufficiently high precision to realize their performance benefits. The present theoretical results, and the findings of our follow-on experimental studies, may have broad applicability to the new and burgeoning fields of high-speed electronics, particularly where sensitivity to small reflections at an input-line/load-bearing interface is of critical importance.

\begin{acknowledgments}
This work was supported by the Army Research Office and the Laboratory for Physical Sciences under EAO221146, EAO241777 and the NIST Quantum Initiative. RPE and Self-Energy, LLC acknowledge grant 70NANB17H033 from the US Department of Commerce, NIST. We have benefited from discussions with D. Pappas, X. Wu, M. Bal, H.-S. Ku, J. Long, R. Lake, L. Ranzani, K. C. Fong, T. Ohki, and B. Abdo.
\end{acknowledgments}

\appendix

\begin{widetext}

\section{\label{appendix:DifferentialEquation}Derivation of the Differential Equation of the Reflection Coefficient of an Impedance Transformer}
Consider an input line and load-bearing transmission line with mismatched characteristic impedances $\mathcal{Z}_1$ and $\mathcal{Z}_2$, respectively. A tapered impedance transformer of length $\ell$ is fabricated within the transmission line to address the mismatch, as in Fig.~\ref{fig1}(a). Each of these three components is modeled as a ladder-type transmission line with unit-length series inductance $L(x)$, series resistance $R(x)$, shunt capacitance $C(x)$, and shunt conductance $G(x)$, comprising a ladder rung at $x$ extending over the infinitesimal length $dx$, as in Fig.~\ref{fig1}(b). Within the input line (interior of the transmission line) $L(x)$, $R(x)$, $C(x)$, and $G(x)$ are all constant and denoted by subscript 1 (2), whereas in the transformer these quantities vary with position $x$, $0\le x\le \ell$. Voltage $V(x,t)$ and current $I(x,t)$ at $x$ satisfy transmission-line equations given by
\begin{subequations} \label{eq:telegrapherEquations}
\begin{eqnarray}
\frac{\partial }{{\partial x}}I(x,t) + C(x)\frac{\partial }{{\partial t}}V(x,t) + G(x) V(x,t) = 0 , \label{eq:telegrapherEquations-1} \\
\frac{\partial }{{\partial x}}V(x,t) + L(x)\frac{\partial }{{\partial t}}I(x,t) + R(x) I(x,t) = 0 . \label{eq:telegrapherEquations-2}
\end{eqnarray}
\end{subequations}
These equations also apply in the input line (interior of the transmission line), except that $L(x)$, $R(x)$, $C(x)$, and $G(x)$ are replaced by $L_1$, $R_1$, $C_1$, and $G_1$ ($L_2$, $R_2$, $C_2$, and $G_2$).

\subsection{Region Before the Taper}
Within the region before the taper, i.e., $x<0$ in Fig.~\ref{fig1}(a), for a traveling-wave of frequency $\omega$, the solution of the input line is of the form
\begin{subequations} \label{eq:formBeforeTaper}
\begin{eqnarray}
I(x,t) = I_1(x,\omega) e^{i\omega t} + I_1(x,\omega)^* e^{-i\omega t} , \label{eq:formBeforeTaper-1} \\
V(x,t) = V_1(x,\omega) e^{i\omega t} + V_1(x,\omega)^* e^{-i\omega t} . \label{eq:formBeforeTaper-2} 
\end{eqnarray}
\end{subequations}
Substituting this into the input-line form of Eqs.~(\ref{eq:telegrapherEquations}) we obtain
\begin{subequations}
\begin{eqnarray}
\frac{\partial }{\partial x}I_1(x,\omega) + Y_1(\omega) V_1(x,\omega) = 0 , \\
\frac{\partial }{\partial x}V_1(x,\omega) + Z_1(\omega) I_1(x,\omega) = 0 ,
\end{eqnarray}
\end{subequations}
where we have defined $Z_1(\omega)=R_1+i\omega L_1$ to be a line impedance per unit length and $Y_1(\omega)=G_1+i\omega C_1$ is a shunt admittance per unit length. If we assume amplitudes with spatial dependence of the form $I_1(x,\omega)=A_1(\omega)\exp{[\gamma(\omega) x]}$ and $V_1(x,\omega)=B_1(\omega)\exp{[\gamma(\omega) x]}$ then
\begin{subequations}
\begin{eqnarray}
\gamma(\omega) A_1(\omega) + Y_1(\omega) B_1(\omega) = 0 , \\
Z_1(\omega) A_1(\omega) + \gamma(\omega) B_1(\omega) = 0 ,
\end{eqnarray}
\end{subequations}
such that a non-trivial solution of $A_1(\omega)$ and $B_1(\omega)$ requires $\gamma(\omega)=\pm \gamma_1(\omega)$, where $\gamma_1(\omega)=\sqrt{Z_1(\omega)Y_1(\omega)}$. So we have two solutions we may express as a superposition, viz.
\begin{subequations}
\begin{eqnarray}
I_1(x,\omega) = A^{(+)}_1(\omega) e^{-\gamma_1(\omega) x} + A^{(-)}_1(\omega) e^{\gamma_1(\omega) x} , \\
V_1(x,\omega) = \mathcal{Z}_1(\omega) \left[ A^{(+)}_1(\omega) e^{-\gamma_1(\omega) x} - A^{(-)}_1(\omega) e^{\gamma_1(\omega) x} \right] ,
\end{eqnarray}
\end{subequations}
where the input-line characteristic impedance is $\mathcal{Z}_1(\omega)=\sqrt{Z_1(\omega)\left/ Y_1(\omega)\right.}$. The amplitude $A^{(+)}_1(\omega)$ ($A^{(-)}_1(\omega)$) is that of a forward (backward) traveling wave. In keeping with convention, we may define the reflection coefficient of the input line as the ratio of backward-traveling voltage amplitude to forward-traveling voltage amplitude, viz. $\rho_1(\omega) = \left[ \mathcal{Z}_1(\omega) A^{(-)}_1(\omega) \right] / \left[ -\mathcal{Z}_1(\omega) A^{(+)}_1(\omega) \right] = -A^{(-)}_1(\omega) / A^{(+)}_1(\omega)$.

\subsection{Region After the Taper}
In the interior of the transmission line, after the taper, i.e., $x>\ell$, the solution follows similarly to the region before the taper, viz.
\begin{subequations} \label{eq:formAfterTaper}
\begin{eqnarray}
I(x,t) = I_2(x,\omega) e^{i\omega t} + I_2(x,\omega)^* e^{-i\omega t} , \\
V(x,t) = V_2(x,\omega) e^{i\omega t} + V_2(x,\omega)^* e^{-i\omega t} .
\end{eqnarray}
\end{subequations}
However, in this case there is no backward-traveling component; one has instead
\begin{subequations}
\begin{eqnarray}
I_2(x,\omega) = A^{(+)}_2(\omega) e^{-\gamma_2(\omega) \left( x - \ell \right)} , \\
V_2(x,\omega) = \mathcal{Z}_2(\omega) A^{(+)}_2(\omega) e^{-\gamma_2(\omega) \left( x - \ell \right)} ,
\end{eqnarray}
\end{subequations}
where $Z_2(\omega)=R_2+i\omega L_2$, $Y_2(\omega)=G_2+i\omega C_2$, and also $\mathcal{Z}_2(\omega)=\sqrt{Z_2(\omega)\left/ Y_2(\omega)\right.}$, $\gamma_2(\omega)=\sqrt{Z_2(\omega)Y_2(\omega)}$. In the limit $\ell\rightarrow 0$ then Eq.~(\ref{eq:formBeforeTaper}) matches to Eq.~(\ref{eq:formAfterTaper}) at $x=0$, resulting in
\begin{subequations}
\begin{eqnarray}
A^{(+)}_1(\omega) + A^{(-)}_1(\omega) = A^{(+)}_2(\omega) , \\
\mathcal{Z}_1(\omega) \left[ A^{(+)}_1(\omega) - A^{(-)}_1(\omega) \right] = \mathcal{Z}_2(\omega) A^{(+)}_2(\omega) .
\end{eqnarray}
\end{subequations}
With $\rho_1(\omega)=-A^{(-)}_1(\omega)/A^{(+)}_1(\omega)$, the above equations imply
\begin{equation}
\rho_1(\omega) = \frac{\mathcal{Z}_2(\omega) - \mathcal{Z}_1(\omega)}{\mathcal{Z}_2(\omega) + \mathcal{Z}_1(\omega)} ,
\end{equation}
which is $\rho_1(\omega)$ in the absence of a transformer.

\subsection{Region of the Taper}
The region of the taper corresponds to $0<x<\ell$. Since here the unit-length series impedance $Z(x,\omega)=R(x)+i\omega L(x)$ and shunt admittance $Y(x,\omega)=G(x)+i\omega C(x)$ vary with distance $x$ we segment the taper into subregions small enough that these quantities may be considered constant in each subregion. We then solve for voltage and current in each subregion, as we did for the regions outside the taper, with the caveat that we have to join solutions of the subregions together to obtain the full solution across the length of the taper. Once we accomplish this we can join this full solution of $0<x<\ell$ to that of $x<0$ and $x>\ell$.

Specifically, we segment the taper into $N$ subregions where the n-th subregion corresponds to interval $x_{n-1}<x<x_n$, of length $\Delta x = \ell/N$, i.e., $x_n=x_{n-1}+\Delta x=\ell n / N$, with $x_0\equiv 0$ and $x_N\equiv \ell$. If $\Delta x$ is sufficiently narrow then constituents of impedance and admittance of the n-th subregion may be denoted by constant values $L^{(n)}$, $R^{(n)}$, $C^{(n)}$, and $G^{(n)}$, such that $Z^{(n)}(\omega)=R^{(n)}+i\omega L^{(n)}$ and $Y^{(n)}(\omega)=G^{(n)}+i\omega C^{(n)}$. Then the equations governing the voltage and current of the n-th subregion are of the same form as those before the taper, i.e., $x_{n-1} < x < x_n$, viz.
\begin{subequations} \label{eq:formOfTheTaper}
\begin{eqnarray}
I(x,t) = I^{(n)}(x,\omega) e^{i\omega t} + I^{(n)}(x,\omega)^* e^{-i\omega t} , \label{eq:formOfTheTaper-1} \\
V(x,t) = V^{(n)}(x,\omega) e^{i\omega t} + V^{(n)}(x,\omega)^* e^{-i\omega t} , \label{eq:formOfTheTaper-2}
\end{eqnarray}
\end{subequations}
where we may write
\begin{subequations}
\begin{eqnarray}
I^{(n)}(x,\omega) = A^{(n,+)}(\omega) e^{-\gamma^{(n)}(\omega) \left( x - x_{n-1} \right)} 
+ A^{(n,-)}(\omega) e^{\gamma^{(n)}(\omega) \left( x - x_{n-1} \right)} , \\
V^{(n)}(x,\omega) = \mathcal{Z}^{(n)}(\omega) \left[ A^{(n,+)}(\omega) e^{-\gamma^{(n)}(\omega) \left( x - x_{n-1} \right)} 
- A^{(n,-)}(\omega) e^{\gamma^{(n)}(\omega) \left( x - x_{n-1} \right)} \right] ,
\end{eqnarray}
\end{subequations}
where $\mathcal{Z}^{(n)}(\omega) = \sqrt{Z^{(n)}(\omega) / Y^{(n)}(\omega)}$ and $\gamma^{(n)}(\omega) = \sqrt{Z^{(n)}(\omega)Y^{(n)}(\omega)}$.

At $x=x_n$ we may match the solution of $x_{n-1} < x < x_n$ to that of $x_n < x < x_{n+1}$, viz.
\begin{gather}
A^{(n,+)}(\omega) e^{-\gamma^{(n)}(\omega) \Delta x} + A^{(n,-)}(\omega) e^{\gamma^{(n)}(\omega) \Delta x} = A^{(n+1,+)}(\omega) + A^{(n+1,-)}(\omega) ,  \label{eq:I-n} \\
\mathcal{Z}^{(n)}(\omega) \left[ A^{(n,+)}(\omega) e^{-\gamma^{(n)}(\omega) \Delta x} - A^{(n,-)}(\omega) e^{\gamma^{(n)}(\omega) \Delta x} \right] = 
\mathcal{Z}^{(n+1)}(\omega) \left[ A^{(n+1,+)}(\omega) - A^{(n+1,-)}(\omega) \right] .  \label{eq:V-n}
\end{gather}
As $\Delta x\rightarrow 0$ one passes to the continuum limit wherein
\begin{gather}
A^{(n,\pm)}(\omega) \rightarrow A^{(\pm)}(x,\omega) , \;
A^{(n+1,\pm)}(\omega) \rightarrow A^{(\pm)}(x,\omega) + \Delta x \frac{\partial}{\partial x} A^{(\pm)}(x,\omega) , \nonumber \\
e^{\pm \gamma^{(n)}(\omega) \Delta x} \rightarrow 1 \pm \gamma(x,\omega) \Delta x , \;
\mathcal{Z}^{(n)}(\omega) \rightarrow \mathcal{Z}(x,\omega) , \;
\mathcal{Z}^{(n+1)}(\omega) \rightarrow \mathcal{Z}(x,\omega) + \Delta x \frac{\partial}{\partial x} \mathcal{Z}(x,\omega) .
\end{gather}
Applying these limiting results to the two boundary equations at $x=x_n$, then as $\Delta x\rightarrow 0$ we find
\begin{equation}
-\gamma(x,\omega) \left[ A^{(+)}(x,\omega) - A^{(-)}(x,\omega) \right] = 
\frac{\partial}{\partial x} A^{(+)}(x,\omega) + \frac{\partial}{\partial x} A^{(-)}(x,\omega) ,
\end{equation}
\begin{multline}
-\gamma(x,\omega) \mathcal{Z}(x,\omega) \left[ A^{(+)}(x,\omega) + A^{(-)}(x,\omega) \right] = \\
\mathcal{Z}(x,\omega) \left[ \frac{\partial}{\partial x} A^{(+)}(x,\omega) - \frac{\partial}{\partial x} A^{(-)}(x,\omega) \right] 
+ \left[ \frac{\partial}{\partial x} \mathcal{Z}(x,\omega) \right] \left[ A^{(+)}(x,\omega) - A^{(-)}(x,\omega) \right] .
\end{multline}
Solving for the derivatives of the amplitudes in these two equations yields
\begin{equation} \label{eq:amplitudes-1}
\frac{\partial}{\partial x} A^{(+)}(x,\omega) = -\gamma(x,\omega) A^{(+)}(x,\omega) 
- \frac{1}{2} \left[ \frac{\partial}{\partial x} \log{\mathcal{Z}(x,\omega)} \right] \left[ A^{(+)}(x,\omega) - A^{(-)}(x,\omega) \right] ,
\end{equation}
\begin{equation} \label{eq:amplitudes-2}
\frac{\partial}{\partial x} A^{(-)}(x,\omega) = \gamma(x,\omega) A^{(-)}(x,\omega) 
+ \frac{1}{2} \left[ \frac{\partial}{\partial x} \log{\mathcal{Z}(x,\omega)} \right] \left[ A^{(+)}(x,\omega) - A^{(-)}(x,\omega) \right] .
\end{equation}
These are the first-order differential equations governing the solution of the amplitudes of the taper region, $0<x<\ell$.

As we did for the region before the taper, we may define a reflection coefficient for a position $x$ within the taper as $\rho(x,\omega)=-A^{(-)}(x, \omega)/A^{(+)}(x, \omega)$, which has a derivative with respect to $x$ given by
\begin{equation} \label{eq:rho-diffeq}
\frac{\partial}{\partial x} \rho(x,\omega) = \frac{A^{(-)}(x, \omega) \frac{\partial}{\partial x} A^{(+)}(x, \omega) 
- A^{(+)}(x, \omega) \frac{\partial}{\partial x} A^{(-)}(x, \omega)}{ A^{(+)}(x, \omega)^2 } .
\end{equation}
If we substitute Eqs.~(\ref{eq:amplitudes-1}) and (\ref{eq:amplitudes-2}) into Eq.~(\ref{eq:rho-diffeq}) we obtain
\begin{equation} \label{eq:rho-diffeq-derived}
\frac{\partial}{\partial x} \rho(x,\omega) = 2 \gamma(x,\omega) \rho(x,\omega)
- \frac{1}{2} \left[ \frac{\partial}{\partial x} \log{\mathcal{Z}(x,\omega)} \right] \left[ 1 - \rho(x,\omega)^2 \right] .
\end{equation}
This is the expression derived by Walker and Wax.\cite{Walker1946} We next consider boundary conditions applicable to this differential equation.

First, at $x=0$, we can equate the current and voltage of Eqs.~(\ref{eq:formBeforeTaper}), corresponding to the region before the taper, to the current and voltage of Eqs.~(\ref{eq:formOfTheTaper}), corresponding to the subregion of $n=1$, just inside the taper. As $\Delta x \rightarrow 0$ the result is
\begin{subequations}
\begin{eqnarray}
A^{(+)}_1(\omega) \left[ 1 - \rho_1(\omega) \right] = A^{(+)}(0,\omega) \left[ 1 - \rho(0,\omega) \right] , \\
\mathcal{Z}_1(\omega) A^{(+)}_1(\omega) \left[ 1 + \rho_1(\omega) \right] = \mathcal{Z}(0,\omega) A^{(+)}(0,\omega) \left[ 1 + \rho(0,\omega) \right] ,
\end{eqnarray}
\end{subequations}
where the reflection coefficient of the input line is $\rho_1(\omega)=-A^{(-)}_1(\omega)/A^{(+)}_1(\omega)$ and, similarly, the reflection coefficient of the taper at $x=0$ is $\rho(0,\omega)=-A^{(-)}(0,\omega)/A^{(+)}(0,\omega)$. If we divide the first equation into the second and solve for $\rho_1(\omega)$ we find
\begin{equation} \label{eq:rho-1}
\rho_1(\omega) = \frac{ \mathcal{Z}(0,\omega) \left[ 1 + \rho(0,\omega) \right] - \mathcal{Z}_1(\omega) \left[ 1 - \rho(0,\omega) \right] }
{ \mathcal{Z}(0,\omega) \left[ 1 + \rho(0,\omega) \right] + \mathcal{Z}_1(\omega) \left[ 1 - \rho(0,\omega) \right] } .
\end{equation}
The reflection coefficient is discontinuous across the physical boundary at $x=0$ unless the two lines on either side of the interface have the same characteristic impedance at $x=0$.

Similarly, at $x=\ell$, we can consider the continuity of current and voltage across this interface using Eqs.~(\ref{eq:formOfTheTaper}), corresponding to the subregion of $n=N$, just inside the taper, and Eqs.~(\ref{eq:formAfterTaper}), corresponding to the region after the taper. As $\Delta x \rightarrow 0$ we find
\begin{subequations}
\begin{eqnarray}
A^{(+)}(\ell,\omega) \left[ 1 - \rho(\ell,\omega) \right] = A^{(+)}_2(\omega) , \\
\mathcal{Z}(\ell,\omega) A^{(+)}(\ell,\omega) \left[ 1 + \rho(\ell,\omega) \right] = \mathcal{Z}_2(\omega) A^{(+)}_2(\omega) .
\end{eqnarray}
\end{subequations}
Again, dividing the first equation into the second and solving for $\rho(\ell,\omega)$, we obtain
\begin{equation} \label{eq:rho-ell}
\rho(\ell,\omega) = \frac{ \mathcal{Z}_2(\omega) - \mathcal{Z}(\ell,\omega) }{ \mathcal{Z}_2(\omega) + \mathcal{Z}(\ell,\omega) } .
\end{equation}
If the interface at $x=\ell$ is not a physical one then  $\mathcal{Z}(\ell,\omega)=\mathcal{Z}_2(\omega)$, which implies $\rho(\ell,\omega)=0$.

\section{\label{appendix:LosslessSolution}Physically Meaningful Solutions of the Lossless Impedance Transformer}
To obtain the solution of the lossless impedance transformer we apply Eqs.~(\ref{eq:lossless}) to Eq.~(\ref{eq:constraint-0}). Separating Eq.~(\ref{eq:constraint-0}) into real and imaginary parts, we may write
\begin{multline} \label{eq:constraint-3}
2 \left| \gamma(0,\omega_o) \right| \left[ \mathcal{Z}_1 - \mathcal{Z}(0,\omega_o) \right] \Re{f(x,\omega_o)} \\
- \left| \gamma(x,\omega_o) \right| \Bigg\{ 4 \left| \gamma(0,\omega_o) \right| \left[ \mathcal{Z}_1 - \mathcal{Z}(0,\omega_o) \right] \int_x^{\ell} \Im{f(x',\omega_o)} \, dx' 
- \left[ \mathcal{Z}_1 + \mathcal{Z}(0,\omega_o) \right] \log{ \frac{\mathcal{Z}(0,\omega_o)}{\mathcal{Z}_2} } \Bigg\} = 0 ,
\end{multline}
\begin{equation} \label{eq:constraint-4}
2 \left| \gamma(0,\omega_o) \right| \left[ \mathcal{Z}_1 - \mathcal{Z}(0,\omega_o) \right] \Bigg\{ \Im{f(x,\omega_o)} + 2 \left| \gamma(x,\omega_o) \right| \int_x^{\ell} \Re{f(x',\omega_o)} \, dx' \Bigg\} = 0 ,
\end{equation}
where $|\gamma(x,\omega_o)|=\omega_o\sqrt{L(x)C(x)}$ and $\mathcal{Z}(x,\omega_o)=\sqrt{(L(x)/C(x)}$. As in our earlier discussion of the more general case, for the lossless transformer a viable choice of $\rho(x,\omega_o)$, i.e., $f(x,\omega_o)$ in Eq.~(\ref{eq:rho-solution}), depends on the existence of a physically meaningful solution of $L(x)$ and $C(x)$. As we shall show, restrictions on the solution will narrow the possible choices for $f(x,\omega_o)$.

For example, consider Eqs.~(\ref{eq:constraint-3}) and (\ref{eq:constraint-4}) in the limit $x\rightarrow 0$, viz.
\begin{equation} \label{eq:0-re} 
\left| \gamma(0,\omega_o) \right| \Bigg\{ 2 \left[ \mathcal{Z}_1 - \mathcal{Z}(0,\omega_o) \right] 
\left[ 1 - 2 \left| \gamma(0,\omega_o) \right| \int_0^{\ell} \Im{f(x',\omega_o)} \, dx' \right] 
+ \left[ \mathcal{Z}_1 + \mathcal{Z}(0,\omega_o) \right] \log{ \frac{\mathcal{Z}(0,\omega_o)}{\mathcal{Z}_2} } \Bigg\} = 0 , 
\end{equation}
\begin{equation} \label{eq:0-im}
{ \left| \gamma(0,\omega_o) \right| }^2 \left[ \mathcal{Z}_1 - \mathcal{Z}(0,\omega_o) \right] \int_0^{\ell} \Re{f(x',\omega_o)} \, dx' = 0 . 
\end{equation}
We see in Eq.~(\ref{eq:0-im}) that the real part of the integral of $f(x,\omega_o)$ must vanish since alternatively (i) $\mathcal{Z}(0,\omega_o)=\mathcal{Z}_1 $, corresponding to a trivial solution of impedance matching, or (ii) $|\gamma(0,\omega_o)|=0$, which implies an infinite group velocity at the materials interface. Thus, inclusive of the boundary conditions of $f(x,\omega_o)$ stated in Eq.~(\ref{eq:rho-solution}), constraints imposed on $f(x,\omega_o)$ are
\begin{equation} \label{eq:choice-1}
f(0,\omega_o) = 1 , \;\;\ f(\ell,\omega_o) = 0 , \;\; \int_0^{\ell} \Re{f(x,\omega_o)} \, dx = 0 .
\end{equation}

As another example, since $|\gamma(0,\omega_o)|\ne 0$, we have from Eq.~(\ref{eq:constraint-4}) that
\begin{equation} \label{eq:LC}
\left| \gamma(x,\omega_o) \right| = \frac{ -\Im{f(x,\omega_o)} }{ 2 \int_x^{\ell} \Re{f(x',\omega_o)} \, dx' } ,
\end{equation}
where, via L'Hospital's rule, we find
\begin{equation}
\left| \gamma(\ell,\omega_o) \right| = \lim\limits_{x\rightarrow\ell} \left| \gamma(x,\omega_o) \right| = 
\frac{1}{2} \lim\limits_{x\rightarrow\ell} \frac{ \Im{\frac{\partial}{\partial x} f(x,\omega_o)} }{ \Re{f(x,\omega_o)} } .
\end{equation}
However, unless $\Im{\partial f(x,\omega_o)/\partial x} =0$ as $x\rightarrow\ell$, the above limit will be unbounded since $\Re{f(\ell,\omega_o)}=0$ via Eq.~(\ref{eq:rho-solution}). Therefore, insisting $\Im{\partial f(x,\omega_o)/\partial x}$ be zero at $x=\ell$, we may use L'Hospital's rule again, viz.
\begin{equation} \label{eq:LC-ell}
\left| \gamma(\ell,\omega_o) \right| = \frac{1}{2} \lim\limits_{x\rightarrow\ell} 
\frac{ \Im{\frac{\partial^2}{\partial x^2} f(x,\omega_o)} }{ \Re{\frac{\partial}{\partial x} f(x,\omega_o)} } .
\end{equation}
Thus, additional constraints imposed on our choice of $f(x,\omega_o)$ are 
\begin{equation} \label{eq:choice-2}
{ \left. \Im{\frac{\partial}{\partial x} f(x,\omega_o)} \right| }_{x = \ell} = 0 , \;\;\;\;\;\; 
{ \left. \Im{\frac{\partial^2}{\partial x^2} f(x,\omega_o)} \right| }_{x=\ell} = 
2 \left| \gamma(\ell,\omega_o) \right| { \left. \Re{\frac{\partial}{\partial x} f(x,\omega_o)} \right| }_{x=\ell} .
\end{equation}

As a last point of observation, we note from Eq.~(\ref{eq:LC}) that
\begin{equation}
\left| \gamma(0,\omega_o) \right| = \lim\limits_{x\rightarrow 0} \left| \gamma(x,\omega_o) \right| = 
\frac{1}{2} \lim\limits_{x\rightarrow 0} \frac{ \Im{\frac{\partial}{\partial x} f(x,\omega_o)} }{ \Re{f(x,\omega_o)} } .
\end{equation}
Since $\Re{f(0,\omega_o)}=1$ this limit requires
\begin{equation} \label{eq:choice-3}
{ \left. \Im{\frac{\partial}{\partial x} f(x,\omega_o)} \right| }_{x=0} \ne 0 ,
\end{equation}
otherwise we would again have $|\gamma(0,\omega_o)|=0$. Thus, Eqs.~(\ref{eq:choice-1}), (\ref{eq:choice-2}), and (\ref{eq:choice-3}) summarize the constraints on our choice of $f(x,\omega_o)$ to ensure a physically meaningful solution of $L(x)$ and $C(x)$, if one does indeed exist.

The task of choosing $f(x,\omega_o)$, subject to the constraints of Eqs.~(\ref{eq:choice-1}), (\ref{eq:choice-2}), and (\ref{eq:choice-3}), is made easier noting these constraints are satisfied if $f(x,\omega_o)$ is written in the form
\begin{equation} \label{eq:choice-for-f}
f(x,\omega_o) = \frac{d}{dx} g(x) + 2 \gamma(\ell,\omega_o) \, g(x) ,
\end{equation}
where the real-valued function $g(x)$, in turn, satisfies the boundary conditions
\begin{equation} \label{eq:g-bc}
g(0) = 0 , \;\;\ g(\ell) = 0 , \;\; { \left. \frac{d}{dx} g(x) \right| }_{x=0} = 1 , \;\; { \left. \frac{d}{dx} g(x) \right| }_{x=\ell} = 0 .
\end{equation}
Thus, the definition of an optimal lossless impedance transformer reduces to choosing a real-valued $g(x)$ that satisfies the boundary conditions of Eq.~(\ref{eq:g-bc}).

We may express the solution for the optimal lossless impedance transformer in terms of $g(x)$. To start, using $f(x,\omega_o)$ of Eq.~(\ref{eq:choice-for-f}), the solution of $|\gamma(x,\omega_o)|$ in Eq.~(\ref{eq:LC}) becomes
\begin{equation} \label{eq:LC-solution}
\left| \gamma(x,\omega_o) \right| = \frac{ -\Im{f(x,\omega_o)} }{ 2 \int_x^{\ell} \Re{f(x')} \, dx' } = 
\left| \gamma(\ell,\omega_o) \right| \left[ \frac{g(x)}{g(x) - g(\ell)} \right] = \left| \gamma(\ell,\omega_o) \right| ,
\end{equation}
since $g(\ell)=0$. Thus $\left| \gamma(x,\omega_o) \right|$ is constant in $x$, along the entire length of the optimal impedance transformer, for any choice of $g(x)$.

Then, making use of Eqs.~(\ref{eq:choice-for-f}) and (\ref{eq:LC-solution}), we have from Eq.~(\ref{eq:0-re}) the result
\begin{equation} \label{eq:solution-0}
2 \left[ \mathcal{Z}_1 - \mathcal{Z}(0,\omega_o) \right] \left[ 1 - 4 { \left| \gamma(\ell,\omega_o) \right| }^2 \int_0^{\ell} g(x') \, dx' \right] 
+ \left[ \mathcal{Z}_1 + \mathcal{Z}(0,\omega_o) \right] \log{ \frac{\mathcal{Z}(0,\omega_o)}{\mathcal{Z}_2} } = 0 ,
\end{equation}
which determines the solution of $\mathcal{Z}(0,\omega_o)$. Similarly, solving for $\mathcal{Z}(x,\omega_o)$ in Eq.~(\ref{eq:constraint-3}), with the aid of Eqs.~(\ref{eq:choice-for-f}), (\ref{eq:LC-solution}), and (\ref{eq:solution-0}), we arrive at
\begin{equation} \label{eq:char-imp-1}
\mathcal{Z}(x,\omega_o) = 
\mathcal{Z}_2 \exp{ \left\{ \left[ \log{ \frac{ \mathcal{Z}(0,\omega_o) }{ \mathcal{Z}_2 } } \right] \left[ \frac{ dg(x) \left/ dx \right. 
- 4 { \left| \gamma(\ell,\omega_o) \right| }^2 \int_x^{\ell} g(x') \, dx' }{ 1 - 4 { \left| \gamma(\ell,\omega_o) \right| }^2 \int_0^{\ell} g(x') \, dx' } \right] \right\} } .
\end{equation}
Thus, once we determine $\mathcal{Z}(0,\omega_o)$ from Eq.~(\ref{eq:solution-0}) we may obtain $\mathcal{Z}(x,\omega_o)$ via Eq.~(\ref{eq:char-imp-1}). The solutions for the inductance and capacitance per unit length then follow as
\begin{equation}
L(x,\omega_o) = \sqrt{L_2 C_2} \, \mathcal{Z}(x,\omega_o) , \;\; C(x,\omega_o) = \frac{ \sqrt{L_2 C_2} }{ \mathcal{Z}(x,\omega_o) } .
\end{equation}

\section{\label{appendix:HighPass}Derivation of the Wide-Band High-Pass Lossless Impedance Transformer}
A $2N$-degree polynomial 
\begin{equation} \label{eq:g-poly-N}
g^{(HP)}(x,N) = 
\ell \Bigg[ \frac{x}{\ell} - \alpha \sum\limits_{n=1}^{N} a^{(N)}_n { \left( \frac{x}{\ell} \right) }^{2N-n} + \alpha { \left( \frac{x}{\ell} \right) }^{2N} \Bigg]
\end{equation}
can be chosen for $g(x)$ to eliminate all terms to order $N$ on the right side of Eq.~(\ref{eq:expansion-hp}). Here, polynomial coefficients $a^{(N)}_1, \dots, a^{(N)}_N$, and scaling factor $\alpha$, are real-valued, and $g^{(HP)}(x,N)$ satisfies the boundary conditions of Eq.~(\ref{eq:g}) via constraints
\begin{equation} \label{eq:g-bc-hp}
\sum\limits_{n=1}^{N} a^{(N)}_n = 1 + \frac{1}{\alpha} \; , \;
\sum\limits_{n=1}^{N} \left( 2N - n \right) a^{(N)}_n = 2N + \frac{1}{\alpha} .
\end{equation}
The second-order term on the right of Eq.~(\ref{eq:expansion-hp}) is removed in the limit $\alpha\rightarrow 0$ when the equation is divided by
\begin{equation}
\int_0^{\ell} g^{(HP)}(x,N) \, dx = 
\left( \frac{1}{2} - \alpha \sum\limits_{n=1}^{N} \frac{ a^{(N)}_n }{ 2N -n + 1 } + \frac{ \alpha }{ 2N + 1 } \right) \ell^2 .
\end{equation}
Higher terms are eliminated by requiring $g^{(n-1)}(0)=0$ and $g^{(n-1)}(\ell)=0$ for $n=3,\dots,N$. The form of the polynomial guarantees $g^{(n-1)}(0)=0$ while $g^{(n-1)}(\ell)=0$ is obtained if
\begin{equation} \label{eq:g-rest}
\sum\limits_{n=1}^{N} \frac{ \left( 2N - n \right) ! }{ \left( 2N - n - m \right) ! } a^{(N)}_n = \frac{ \left( 2N \right) ! }{ \left( 2N - m \right) ! } \; ; 
\;\; m = 2, 3, \dots, N-1 .
\end{equation}
The $N$ coupled linear algebraic Eqs.~(\ref{eq:g-bc-hp}) and (\ref{eq:g-rest}) determine coefficients $a^{(N)}_1, \dots, a^{(N)}_N$ of the polynomial. 

Substituting Eq.~(\ref{eq:g-poly-N}) into Eq.~(\ref{eq:rho1-hp}) for $g(x)$, and taking the limit $\alpha\rightarrow\infty$, we may write
\begin{multline} \label{eq:rho1-N-hp}
\rho^{(HP)}_1(\omega,N) \cong 
-\frac{1}{2} \left( \log{ \frac{\mathcal{Z}_1}{\mathcal{Z}_2} } \right) { \left[ \sum\limits_{n=0}^{N} \frac{ a^{(N)}_n }{ 2N-n+1 } \right] }^{-1} \\
\times \sum\limits_{n=0}^{N} a^{(N)}_n  \left( 2N - n \right) ! { \left( \frac{ \omega_c }{ 2\pi i\omega } \right) }^{2N-n+1}
\left[ 1 - e^{-2\pi i \left( \omega / \omega_c \right)} 
\sum\limits_{m=0}^{2N-n} \frac{ 1 }{ m! } { \left( \frac{ 2\pi i \omega }{ \omega_c } \right) }^m \right] ,
\end{multline}
where we define $a^{(N)}_0=-1$ and note polynomial coefficients $a^{(N)}_1, \dots, a^{(N)}_N$ now satisfy
\begin{equation} \label{eq:poly-coefficients}
\sum\limits_{n=1}^{N} \frac{ \left( 2N - n \right) ! }{ \left( 2N - n - m \right) ! } a^{(N)}_n = \frac{ \left( 2N \right) ! }{ \left( 2N - m \right) ! } \; ; 
\;\; m = 0, 1, \dots, N-1 .
\end{equation}
Including the definition $a^{(N)}_0=-1$, the solution of Eq.~(\ref{eq:poly-coefficients}) is
\begin{equation} \label{eq:pascal-solution}
a^{(N)}_n = { \left( -1 \right) }^{n+1} \frac{ N! }{ n! \left( N -n \right)! } \; ; \;\; n = 0, 1, \dots, N .
\end{equation}
Similarly, substituting Eq.~(\ref{eq:g-poly-N}) into Eq.~(\ref{eq:Z-hp}) for $g(x)$, and taking the limit $\alpha\rightarrow\infty$, the corresponding characteristic impedance is
\begin{equation} \label{eq:Z-N-hp}
\mathcal{Z}^{(HP)}(x,N) = \mathcal{Z}_2 \exp{ \left( \left( \log{\frac{ \mathcal{Z}_1 }{ \mathcal{Z}_2 }} \right) 
\left\{ 1 - { \left[ \sum\limits_{n=0}^{N} \frac{a^{(N)}_n}{2N-n+1} \right] }^{-1}
\sum\limits_{n=0}^{N} \frac{a^{(N)}_n}{2N-n+1} { \left( \frac{x}{\ell} \right) }^{2N-n+1} \right\} \right) } .
\end{equation}
Equations (\ref{eq:rho1-N-hp}), (\ref{eq:pascal-solution}), and (\ref{eq:Z-N-hp}) define the high-pass transformer design of the $2N$-degree polynomial choice for $g(x)$.

Expanding $\exp{\left( 2\pi i\omega/\omega_c \right)}$ in an infinite sum, an alternate expression of Eq.~(\ref{eq:rho1-N-hp}) is 
\begin{equation} \label{eq:rho1-N-hp-alternate}
\rho^{(HP)}_1(\omega,N) \cong 
-\frac{1}{2} \left( \log{ \frac{\mathcal{Z}_1}{\mathcal{Z}_2} } \right) { \left[ \sum\limits_{n=0}^{N} \frac{ a^{(N)}_n }{ 2N-n+1 } \right] }^{-1} 
\sum\limits_{n=0}^{N} a^{(N)}_n \sum\limits_{m=0}^{\infty} 
\frac{ \left( 2N - n \right) ! }{ \left( 2N - n + m + 1 \right) ! } { \left( \frac{ 2\pi i \omega }{ \omega_c } \right) }^{m} e^{-2\pi i \left( \omega / \omega_c \right)} .
\end{equation}
The coefficients of Eq.~(\ref{eq:pascal-solution}) satisfy the summation identities
\begin{subequations}
\begin{eqnarray}
\sum\limits_{n=0}^{N} \frac{ a^{(N)}_n }{ \left( 2N - n + m \right) } = 
{ \left( -1 \right) }^{N+1} \frac{ N! \left( N + m - 1 \right)! }{ \left( 2N + m \right)! } \; ; \;\; m > 0 , \label{eq:sum-3} \\
\sum\limits_{n=0}^{N} \frac{ \left( 2N - n \right) ! }{ \left( 2N - n + m \right) ! } a^{(N)}_n = 
{ \left( -1 \right) }^{N+1} \frac{ N! \left( N + m - 1 \right)! }{ \left( m - 1 \right)! \left( 2N + m \right)! } \; ; \;\; m > 0 \label{eq:sum-4} .
\end{eqnarray}
\end{subequations}
Using Eq.~(\ref{eq:sum-3}) and (\ref{eq:sum-4}) to evaluate the sums over $n$ in Eq.~(\ref{eq:rho1-N-hp-alternate}), we have
\begin{multline}
\rho^{(HP)}_1(\omega,N) \cong 
-\frac{1}{2} \left( \log{ \frac{\mathcal{Z}_1}{\mathcal{Z}_2} } \right) \frac{ \left( 2N+1 \right)! }{ N! }
\sum\limits_{m=0}^{\infty} \frac{ \left( N + m \right)! }{ m! \left( 2N + m + 1 \right) ! } { \left( \frac{ 2\pi i \omega }{ \omega_c } \right) }^{m}
e^{-2\pi i \left( \omega / \omega_c \right)} = \\ 
-\frac{1}{2} \left( \log{ \frac{\mathcal{Z}_1}{\mathcal{Z}_2} } \right) \, M\left( n+1, 2n+2, \frac{2\pi i\omega}{\omega_c} \right)
e^{-2\pi i \left( \omega / \omega_c \right)} ,
\end{multline}
where in the last step we recognize the resulting sum as Kummer's confluent hypergeometric function, $M(a,b,z)$.\cite[p.~504]{AbramowitzAndStegun} Alternatively, $M(n+1,2n+2,,2iz)=\Gamma(3/2+n)e^{iz}(2/z)^n j_n(z)/\sqrt{\pi}$, where $\Gamma(z)$ is the gamma function and $j_n(z)$ is a spherical Bessel function.\cite[p.~509]{AbramowitzAndStegun}. Thus, we have
\begin{equation} \label{rho1-N-hp-below}
\rho^{(HP)}_1(\omega,N) \cong 
-\frac{ \Gamma\left( 3/2+N \right) }{\sqrt{\pi}} \left( \log{ \frac{\mathcal{Z}_1}{\mathcal{Z}_2} } \right) { \left( \frac{2\omega_c}{\pi\omega} \right) }^N 
j_N \left( \pi\omega/\omega_c \right) \, e^{-i\pi \left( \omega / \omega_c \right)} .
\end{equation}
When $N\gg 1$ we may use the asymptotic form of the Bessel function for large order,\cite[p.~365]{AbramowitzAndStegun} such that
\begin{equation}  \label{eq:rho1-poly-lessthan}
\rho^{(HP)}_1(\omega,N) \cong 
-\frac{1}{2} \left( \log{ \frac{\mathcal{Z}_1}{\mathcal{Z}_2} } \right) 
e^{-i\pi \left( \omega / \omega_c \right)} \; ; 
\;\; \frac{\pi\omega}{\omega_c}\ll N \gg 1 .
\end{equation}
Also, by construction, $\rho^{(HP)}_1(\omega,N)\propto 1/\omega^{N+1}$ as $\omega\rightarrow\infty$.

With the aid of Eq.~(\ref{eq:sum-3}), we may write Eq.~(\ref{eq:Z-N-hp}) as
\begin{equation}
\mathcal{Z}^{(HP)}(x,N) = \mathcal{Z}_2 
\exp{ \left\{ \left( \log{\frac{ \mathcal{Z}_1 }{ \mathcal{Z}_2 }} \right) 
\left[ 1 - { \left( -1 \right) }^{N+1} \frac{ \left( 2N + 1 \right)! }{ { \left( N! \right) }^2 } S \left( \frac{x}{\ell} \right) \right] \right\} } ,
\end{equation}
where we have introduced the sum
\begin{equation}
S(z) = \sum\limits_{n=0}^{N} \frac{a^{(N)}_n}{2N-n+1} z^{2N-n+1} .
\end{equation}
Differentiating $S(z)$ with respect to $z$ and making use of Eq.~(\ref{eq:poly-coefficients}), as well as the binomial theorem, we find derivative $S'(z)=-{\left(z^2-z\right)}^N$. Integrating $S'(z)$ from $0$ to $z$ we arrive at the expression
\begin{equation} \label{eq:Z-N-hp-interim}
\mathcal{Z}^{(HP)}(x,N) = \mathcal{Z}_2 \exp{ \left\{ \left( \log{\frac{ \mathcal{Z}_1 }{ \mathcal{Z}_2 }} \right) 
\left[ 1 - I\left( x/\ell;N+1,N+1 \right) \right] \right\} } ,
\end{equation}
where
\begin{equation}
I\left(z;N+1,N+1 \right) = \frac{ \left( 2N + 1 \right)! }{ { \left( N! \right) }^2 } \int_0^z { \left( u - u^2 \right) }^N du 
\end{equation}
is a regularized incomplete beta function.\cite[p.~263]{AbramowitzAndStegun} Since $I(z;a,b)=1-I(1-z;b,a)$ then $I(1/2;a,a)=1/2$ such that $\mathcal{Z}^{(HP)}(\ell/2,N)=\sqrt{\mathcal{Z}_1\mathcal{Z}_2}$ is a fixed point, the same for all $N$, and Eq.~(\ref{eq:Z-N-hp-interim}) may be expressed alternatively as
\begin{equation} \label{eq:Z-N-hp-appendix}
\mathcal{Z}^{(HP)}(x,N) = \mathcal{Z}_2 \exp{ \left[ \left( \log{\frac{ \mathcal{Z}_1 }{ \mathcal{Z}_2 }} \right) 
I\left( 1 - x/\ell;N+1,N+1 \right) \right] } .
\end{equation}
With $0\le z\le 1$, as $N\rightarrow\infty$ we have\cite{Pagurova1995}
\begin{equation}
\lim\limits_{N\rightarrow\infty} I\left(z;N+1,N+1 \right) = \lim\limits_{N\rightarrow\infty} 
\frac{1}{\sqrt{2\pi}} \int_{-\infty}^{3 \sqrt{N} \phi(z)} e^{-u^2/2} du \; ; \;\;
\phi(z) = \frac{ z^{1/3} - {\left( 1 - z \right) }^{1/3} }{ \sqrt{ z^{1/3} + {\left( 1 - z \right) }^{1/3} } } .
\end{equation}
Since $\phi(z)>0$ if $z>1/2$ and $\phi(z)<0$ if $z<1/2$ then
\begin{equation}
\lim\limits_{N\rightarrow\infty} I\left(z;N+1,N+1 \right) =
\Theta \left( z - 1/2 \right) ,
\end{equation}
where $\Theta(z)$ is the Heaviside step function. Therefore, we may write
\begin{equation}
\mathcal{Z}^{(HP)}(x) = \lim\limits_{N\rightarrow\infty} \mathcal{Z}^{(HP)}(x,N) = 
\mathcal{Z}_2 \exp{ \left[ \left( \log{\frac{ \mathcal{Z}_1 }{ \mathcal{Z}_2 }} \right) \Theta \left( \ell/2 - x \right) \right] } .
\end{equation}

\section{\label{appendix:LowPass}Derivation of the Wide-Band Low-Pass Lossless Impedance Transformer}
Analogous to the derivation of the wide-band high-pass lossless transformer, we construct a $(N+3)$-degree polynomial
\begin{equation} \label{eq:g-lp}
g^{(LP)}(x,N) = \ell \left[ \frac{x}{\ell} - \sum\limits_{n=1}^{N+2} a^{(N)}_n { \left( \frac{x}{\ell} \right) }^{n+1} \right] ,
\end{equation}
where $g^{(LP)}(x,N)$ satisfies the boundary conditions of Eq.~(\ref{eq:g}) via constraints
\begin{equation} \label{eq:bc-lp}
\sum\limits_{n=1}^{N+2} \left( n + 1 \right) a^{(N)}_n = 1 \; , \; 
\sum\limits_{n=1}^{N+2} a^{(N)}_n = 1 . 
\end{equation}
To eliminate terms to order $N-1$ in $2\pi\omega/\omega_c$ of Eq.~(\ref{eq:expansion-lp}) we set $G^{(n)}(\ell)=0$ for $n=1,\dots,N$, which introduces $N$ additional constraints involving polynomial coefficients $a^{(N)}_1, \dots, a^{(N)}_{N+2}$. Combining these constraints with those of Eq.~(\ref{eq:bc-lp}) we obtain a set of $N+2$ coupled linear algebraic equations in $a^{(N)}_1, \dots, a^{(N)}_{N+1}$ that we may express as
\begin{equation} \label{eq:constraints-lp}
\sum\limits_{n=1}^{N+2} \frac{ \left( n + 1 \right)! }{ \left( n + m \right)! } a^{(N)}_n = \frac{1}{m!} \; ; \;\; m = 0,1, \dots, N+1 . 
\end{equation}
The solution is
\begin{equation} \label{eq:solution-lp}
a^{(N)}_n = { \left( -1 \right) }^{n+1} \frac{ N + 2 }{ n! \left( n + 1 \right)! } \, \frac{ \left( N + n + 1 \right)! }{ \left( N - n + 2 \right)! } \; ; 
\;\; n = 0,1,\dots,N+2 ,
\end{equation}
where we have included $a^{(N)}_0=-1$ for completeness.

Substituting Eq.~(\ref{eq:g-lp}) into Eq.~(\ref{eq:rho1-lp}) for $g(x)$, we obtain
\begin{equation} \label{eq:rho1-lp-sum}
\rho^{(LP)}_1(\omega,N) \cong
-\frac{1}{2} \left[ \log{ \frac{\mathcal{Z}(0,0)}{\mathcal{Z}_2} } \right] \sum\limits_{n=0}^{N+2} a^{(N)}_n \left( n+1 \right)! 
{ \left( \frac{ \omega_c }{ 2\pi i \omega } \right) }^n
\left[ 1 - e^{-2\pi i \left( \omega \left/ \omega_c \right. \right) } \sum\limits_{m=0}^{n+1} \frac{1}{m!} { \left( \frac{ 2\pi i \omega }{ \omega_c } \right) }^m \right] .
\end{equation}
Expanding $\exp{\left( 2\pi i\omega/\omega_c \right)}$ in an infinite sum, an alternate expression is
\begin{equation} \label{eq:rho1-N-lp-interim}
\rho^{(LP)}_1(\omega,N) \cong
-\frac{1}{2} \left[ \log{ \frac{\mathcal{Z}(0,0)}{\mathcal{Z}_2} } \right] 
\sum\limits_{m=0}^{\infty} \left[ \sum\limits_{n=0}^{N+2} a^{(N)}_n \frac{ \left( n+1 \right)! }{ \left( m+n+2 \right)! } \right] 
{ \left( \frac{ 2\pi i \omega }{ \omega_c } \right) }^{m+2} e^{-2\pi i \left( \omega \left/ \omega_c \right. \right) } ,
\end{equation}
where we note that the sum over $n$ may be written in closed form as
\begin{equation}
\sum\limits_{n=0}^{N+2} a^{(N)}_n \frac{ \left( n+1 \right)! }{ \left( m+n+2 \right)! } = \left\{
\begin{array}{ll}
0 & ; \; m<N \\
\frac{ -\left( m + 2 \right)! }{ \left( m - N \right)! \, \left( m + N + 4 \right)! } & ; \; m \ge N
\end{array} \right. .
\end{equation}
Applying this last result to Eq.~(\ref{eq:rho1-N-lp-interim}), and then shifting the sum over $m$ by $N$, we obtain
\begin{equation}
\rho^{(LP)}_1(\omega,N) \cong
\frac{1}{2} \left[ \log{ \frac{\mathcal{Z}(0,0)}{\mathcal{Z}_2} } \right] \sum\limits_{m=0}^{\infty} 
\frac{ \left( N + m + 2 \right)! }{ m! \, \left( 2N + m + 4 \right)! } 
{ \left( \frac{ 2\pi i \omega }{ \omega_c } \right) }^{N+m+2} e^{-2\pi i \left( \omega \left/ \omega_c \right. \right) } .
\end{equation}
Then, similar to the high-pass case, we recognize this sum to be related to Kummer's confluent hypergeometric function, $M(a,b,z)$,\cite[p.~504]{AbramowitzAndStegun} viz.
\begin{equation}
\rho^{(LP)}_1(\omega,N) \cong
\frac{1}{2} \frac{ \left( N + 2 \right)! }{ \left( 2N + 4 \right)! } \left[ \log{ \frac{\mathcal{Z}(0,0)}{\mathcal{Z}_2} } \right] 
{ \left( \frac{ 2\pi i \omega }{ \omega_c } \right) }^{N+2} M \left( N + 3, 2N + 5 , \frac{ 2\pi i \omega }{ \omega_c } \right) 
e^{-2\pi i \left( \omega \left/ \omega_c \right. \right) } .
\end{equation}
By construction, as $\omega\rightarrow 0$ we have $\rho^{(LP)}_1(\omega,N)\propto \omega^{N+2}$. Conversely, as $\omega\rightarrow\infty$ then $M \left( N + 3, 2N + 5 , \frac{ 2\pi i \omega }{ \omega_c } \right)\rightarrow \left[\left( 2N+4\right)!/\left( N+2 \right)! \right] {\left[ \omega_c/\left(2\pi\omega\right) \right]}^{N+2} \exp{\left( 2\pi i \omega/\omega_c \right)}$,\cite[p.~508]{AbramowitzAndStegun} so for fixed $N$ we have
\begin{equation} \label{eq:rho1-N-lp-large-w}
\lim\limits_{\omega\rightarrow\infty} \rho^{(LP)}_1(\omega,N) \cong \frac{1}{2} \log{ \frac{\mathcal{Z}(0,0)}{\mathcal{Z}_2} } .
\end{equation}

The characteristic impedance $\mathcal{Z}^{(LP)}(x,N)$ corresponding to the design function $g^{(LP)}(x,N)$ is obtained by substituting Eq.~(\ref{eq:g-lp}) into Eq.~(\ref{eq:Z-lp}) for $g(x)$. Within the resulting expression the derivative with respect to $x$ of $g^{(LP)}(x,N)$ is the Gauss series representation of the ordinary hypergeometric function $F(-N-2,N+2;1;x/\ell)={_2F_1}(-N-2,N+2;1;x/\ell)$.\cite[p.~556]{AbramowitzAndStegun} Alternatively, ${_2F_1}(-N-2,N+2;1;x/\ell)=P^{(0,-1)}_{n+2}(1-2x/\ell)$, where $P^{(0,-1)}_{N+2}(1-2x/\ell)$ is a Jacobi polynomial of order $N+2$.\cite[p.~561]{AbramowitzAndStegun} Hence, we have
\begin{equation}
\frac{d}{dx} g^{(LP)}(x,N) = -\sum\limits_{n=0}^{N+2} a^{(N)}_n \left( n+1 \right) { \left( \frac{x}{\ell} \right) }^n = P^{(0,-1)}_{N+2}(1-2x/\ell) ,
\end{equation} 
where $a^{(N)}_n$ is given by Eq.~(\ref{eq:solution-lp}). Therefore, the characteristic impedance may be written as
\begin{equation}
\mathcal{Z}^{(LP)}(x,N) = \mathcal{Z}_2 \exp{ \left\{ \left[ \log{ \frac{ \mathcal{Z}(0,0) }{ \mathcal{Z}_2 } } \right] P^{(0,-1)}_{N+2}(1-2x/\ell) \right\} } .
\end{equation}
As $N\rightarrow\infty$ we may use the Darboux formula for the large-order expansion of Jacobi polynomials.\cite{Szego1975} Letting $\varphi(x) = \arccos{ \left( 1-2x/\ell \right)}$, this gives
\begin{equation} \label{eq:darboux}
\mathcal{Z}^{(LP)}(x) = \lim\limits_{N\rightarrow\infty} \mathcal{Z}^{(LP)}(x,N) = 
\mathcal{Z}_2 \lim\limits_{N\rightarrow\infty} \exp{ \left\{ \left[ \log{ \frac{ \mathcal{Z}(0,0) }{ \mathcal{Z}_2 } } \right] 
\frac{\cos{ \left[ N \varphi(x) - \pi/4 \right] }}{ \sqrt{ \pi N \tan{ \left[ \varphi(x)/2 \right] } } } \right\} }
= 0 \; ; \;\; x > 0 ,
\end{equation}
where $\mathcal{Z}^{(LP)}(0)=\mathcal{Z}(0,0)$.

\end{widetext}

\bibliography{bibliography}

\end{document}